\documentclass[a4paper,11pt]{article}

\usepackage{jcappub} 
\usepackage{bm}
\usepackage{subcaption}
\usepackage{amsmath}

\def\la{\langle}
\def\ra{\rangle}

\def\bkone{\mathbf k_1}
\def\bktwo{\mathbf k_2}
\def\bkthree{\mathbf k_3}

\title{\boldmath Multi-field effects in a simple extension of $R^2$ inflation}

\author[a]{Taro Mori,}
\author[a,b]{Kazunori Kohri}
\author[b]{and Jonathan White}

\affiliation[a]{SOKENDAI (The Graduate University for Advanced Studies), Oho 1-1, Tsukuba 305-0801, Ibaraki, Japan}
\affiliation[b]{Theory Center, IPNS, KEK, Oho 1-1, Tsukuba 305-0801, Ibaraki, Japan}

\emailAdd{moritaro@post.kek.jp}
\emailAdd{kohri@post.kek.jp}
\emailAdd{jwhite@post.kek.jp}

\abstract{
  We consider inflation in the system containing a Ricci scalar squared
  term and a canonical scalar field with quadratic mass term.  In the
  Einstein frame this model takes the form of a two-field inflation
  model with a curved field space, and under the slow-roll approximation
  contains four free parameters corresponding to the masses of the two
  fields and their initial positions.  We investigate how the
  inflationary dynamics and predictions for the primordial curvature
  perturbation depend on these four parameters.  Our analysis is based
  on the $\delta N$ formalism, which allows us to determine predictions
  for the non-Gaussianity of the curvature perturbation as well as for
  quantities relating to its power spectrum.  Depending on the choice of
  parameters, we find predictions that range from those of $R^2$
  inflation to those of quadratic chaotic inflation, with the
  non-Gaussianity of the curvature perturbation always remaining small.
  Using our results we are able to put constraints on the masses of the
  two fields.
}

\begin{document}
\maketitle
\flushbottom

%
%
\section{Introduction}

Recent cosmic microwave background (CMB) observations are in good
agreement with the predictions of inflation, and the data is now so
precise that it can be used to constrain individual models of
inflation~\cite{Ade:2015lrj}.  While observations are still perfectly
consistent with single-field inflation, in the context of high-energy
particle physics theories there is strong motivation to consider
multi-field models.  For instance, when compactifying superstring theory
or supergravity on to four dimensions, many scalar/pseudo-scalar fields
usually appear, such as moduli and axions.  It is thus important to
determine the observable consequences of multi-field inflation models
and how they, and the theories in which they are embedded, can be
constrained by current and future observations.  In relation to this
there are perhaps two key features that distinguish multi-field models
from single field models.  The first is that the curvature perturbation
on constant density slices, $\zeta$, is not necessarily conserved on
super-horizon scales, and the second is that its statistical
distribution may deviate from a Gaussian one.  In the case of
single-field inflation, Maldacena's consistency relation dictates that
the non-Gaussianity of $\zeta$ in the squeezed limit should be
unobservably small, which is a consequence of the fact that $\zeta$ is
conserved on super-horizon scales in single-field inflation
models~\cite{Maldacena:2002vr}.  This suggests that if a relatively
large non-Gaussianity were to be observed, this would be a strong
indication that multiple fields were present during inflation.  Even if
not observed, however, it is still important to determine the
implications of this for multi-field models.  While current constraints
on non-Gaussianity from the CMB are relatively weak, future large scale
structure surveys promise to improve these constraints
considerably, see e.g.~\cite{Yamauchi:2016ypt}.  In light of the above, it is
clear that in looking to test any multi-field model of inflation one
will need to be able to calculate how the curvature perturbation evolves
on super-horizon scales and how much its statistical distribution
deviates from a Gaussian one.

In this work we consider a simple multi-field extension of so-called
$R^2$ inflation, also sometimes referred to as Starobinsky inflation
\cite{Starobinsky:1980te}.  In its Jordan frame representation, the
original model consists of a modified gravity sector containing a term
proportional to $R^2$, and the inflationary predictions are in very good
agreement with observations \cite{Ade:2015lrj}.  In trying to embed this
model in a more fundamental framework such as a supergravity, however,
it is natural to expect the appearance of additional scalar degrees of
freedom (see e.g.  \cite{Kallosh:2013lkr}) and an important question is
then how much the inflationary predictions are affected by these
additional degrees of freedom.  As a toy model, here we consider adding
a canonical scalar field with quadratic mass term to the original Jordan
frame action, which is the same model as considered in
\cite{vandeBruck:2015xpa}.  Similar models have also been considered in
\cite{Ellis:2014gxa,Ellis:2014opa}.  Re-writing the model as a
scalar-tensor theory of gravity plus additional scalar, and transforming
to the Einstein frame, this model takes the form of a two-field
inflation model with a non-flat field space.  One of the fields, often
referred to as the scalaron, corresponds to the additional scalar degree
of freedom associated with the $R^2$ term in the original action, and
the second is simply the field we have introduced by hand.  In addition
to the non-flat field space, the potential in the Einstein frame also
contains interactions between the two fields.

In analyzing the inflationary predictions of this model we make use of
the separate Universe approach and $\delta N$
formalism~\cite{Sasaki:1995aw,Sasaki:1998ug,Wands:2000dp,Lyth:2004gb,Lyth:2005fi,Sugiyama:2012tj}.
Due to the non-flat field space and interaction terms in the potential,
it is not possible to calculate $\delta N (= \zeta)$ analytically, and
so we rely on numerical calculations.  Making the slow-roll
approximation, such that only the initial field positions need to be
specified in solving the inflationary dynamics, the model essentially
contains four parameters: the masses of the two fields and their initial
positions.  We explore how the inflationary dynamics and predictions
for the correlation functions of $\zeta$ depend on these four
parameters, and using current observational data we put constraints on
the masses of the two fields.

The paper is organised as follows. In Sec.~\ref{setup}, we explain the
concrete set-up of our model and present the background field
equations. In Sec.~\ref{PSBS} we then review the separate Universe
approach and $\delta N$ formalism, which is used to determine the two-
and three-point correlation functions of the curvature perturbation. In
Sec.~\ref{Num} we briefly describe our numerical method and present the
results of our analysis.  Our findings are then summarised in
Sec.~\ref{summary}.

%
%
\section{Set-up and background equations}
\label{setup}

The model we consider contains an $R^2$ term and an additional scalar
field $\chi$ with a canonical kinetic term. We further assume that the
potential for the $\chi$ field is a simple quadratic.  The action of
this model is thus given by
\begin{equation}
 S_{J} = \int d^{4}x \sqrt{-\tilde g} \bigg[ \frac{M^2_{pl}}{2} \tilde R + \frac{\mu}{2} \tilde R^2 \bigg] +
  \int d^{4}x \sqrt{-\tilde g}\bigg[ - \frac{1}{2} \tilde g^{\mu \nu}\partial_{\mu}\chi \partial_{\nu}\chi
  - \frac{1}{2} m_\chi^2 \chi^{2}\bigg].
  \label{eqR_x1}
\end{equation}
Here the subscript $J$ denotes the Jordan frame, $\tilde g_{\mu\nu}$ is
the Jordan frame metric, $\tilde R$ is the Ricci scalar constructed from
$\tilde g_{\mu\nu}$ and its derivatives, $M_{pl}=1/\sqrt{8\pi G}$ is the
reduced Planck mass, where $G$ is Newton's gravitational constant, and
$\mu$ is a dimensionless parameter.  In analyzing the above model it is
useful to re-write it as a model containing two scalar fields and a
canonical Einstein-Hilbert term, which can be achieved as follows, see
e.g.  \cite{DeFelice:2010aj}.  First we introduce the auxiliary field
$\varphi$, and consider the action
\begin{align}
 S_{J\,{\rm Grav}} = \frac{M_{pl}^2}{2}\int d^4 x\sqrt{-\tilde
 g}\left(f(\varphi) +
 f_{,\varphi}(\varphi)(R-\varphi)\right),\label{auxlag}
\end{align}
where $f(\varphi) = \varphi + \mu \varphi^2/M_{pl}^2$ and $f_{,\varphi}
 = df/d\varphi$.  Minimizing this action with respect to $\varphi$ gives
 the constraint
\begin{align}
 \frac{2\mu}{M_{pl}^2}(R-\varphi) = 0, 
\end{align}  
which for non-zero $\mu$ gives $\varphi = R$.  On substituting $\varphi
 = R$ into \eqref{auxlag} we recover the gravitational part of
 \eqref{eqR_x1}, which confirms the equivalence of these two actions.
 Next we introduce $e^{2\alpha\phi} = 1 + 2\mu \varphi/M_{pl}^2$ with
 $\alpha=1/(\sqrt{6}M_{pl})$, such that \eqref{auxlag} takes the form
\begin{align}
S_{J\,{\rm Grav}} = \int d^4 x \sqrt{-\tilde
 g}\left(\frac{M_{pl}^2}{2}e^{2\alpha\phi}\tilde R
 - \tilde V(\phi)\right), \qquad \tilde V(\phi) = \frac{M_{pl}^4}{8\mu}\left(e^{2\alpha\phi}-1\right)^2.      
\end{align}
Thus we have re-written the gravitational part of the action given in
eq.~\eqref{eqR_x1} as a scalar-tensor theory with a non-minimal coupling
between the scalar field $\phi$ and gravity.  Note, however, that there
is no kinetic term for the $\phi$ field in this representation.  Finally
we make a conformal transformation of the metric, expressing the Jordan
frame metric $\tilde g_{\mu\nu}$ in terms of the so-called Einstein
frame metric $g_{\mu\nu}$ as
\begin{equation}
 g_{\mu\nu} = \Omega^{2} \tilde g_{\mu\nu}, \ \ \ \  {\rm where} \ \ \ \
  \Omega^{2} = e^{2\alpha \phi}.
\end{equation}
On doing so we find that the total action takes the form 
\begin{equation}
 S_E = \int d^{4}x \sqrt{- g} \bigg[ \frac{M^2_{pl}}{2}  R - \frac{ g^{\mu \nu}}{2} ( \partial _{\mu} \phi)( \partial_{\nu} \phi)
  -\frac{1}{2}  g^{\mu \nu}e^{-2\alpha \phi}(\partial_{\mu}\chi)( \partial_{\nu}\chi) - V(\phi,\chi) \bigg],
  \label{Eaction}
\end{equation}
with
\begin{equation}
  V(\phi,\chi) = \frac{3}{4}m^2_\phi M_{pl}^2(1-e^{-2\alpha \phi})^2 + \frac{1}{2} m_\chi^2 e^{-4\alpha \phi} \chi^{2}.
  \label{pot}
\end{equation}
Here we introduced $m^2_\phi=M^2_{pl}/(6\mu)$ and the subscript $E$
denotes the Einstein frame.  The label `Einstein frame' is appropriate
given that the gravity part of the action now takes the canonical
Einstein-Hilbert form, which is somewhat easier to analyze than the
gravity sector of the original action \eqref{eqR_x1}.  Note, however,
that reducing the gravity sector to the canonical Einstein-Hilbert form
has come at the cost of introducing the additional scalar degree of
freedom $\phi$ --- often referred to as the scalaron --- and interaction
terms between the two fields $\phi$ and $\chi$, which appear in the
second term of the potential and in the kinetic term of $\chi$.

Using a more abstract notation, the action (\ref{Eaction}) can be
re-written in the form of a non-linear sigma model as
\begin{equation}
 S_E = \int d^4 x \sqrt{-g} \left[ \frac{M^2_{pl}}{2}  R
     - \frac{1}{2} {\cal G}_{IJ}  g^{\mu\nu} \partial_\mu \phi^I \partial_\nu \phi^J - V (\phi^I) \right].
  \label{Eaction2}
\end{equation}
In our case the Latin indices $I$ and $J$ take on the values $\phi$ and
$\chi$, with $\phi^\phi = \phi$ and $\phi^\chi = \chi$.  ${\cal G}_{IJ}$
is interpreted as the metric on field space, and in our case the
components are given as
\begin{equation}
 {\cal G}_{\phi\phi}=1, \ \ \ {\cal G}_{\chi\chi}=e^{-2\alpha \phi},
  \ \ \ {\cal G}_{\phi\chi}={\cal G}_{\chi\phi}=0.
\end{equation}
Varying the action (\ref{Eaction2}) with respect to $g_{\mu\nu}$,
assuming a Friedmann-Lemaitre-Robertson-Walker (FLRW) metric of the form
$g_{\mu\nu} = {\rm diag}(-1,a^2(t),a^2(t),a^2(t))$ and taking the scalar
fields to be homogeneous, namely $\phi^I = \phi^I(t)$, we obtain the
Friedmann equation
\begin{eqnarray}
 H^2 &=& \frac{1}{3M^2_{pl}} \bigg[ \frac{1}{2}{\cal G}_{IJ}\dot{\phi}^I\dot{\phi}^J + V(\phi^I) \bigg], \label{frdmn}
\end{eqnarray}
and the continuity equation 
\begin{eqnarray}
 \dot{H}&=& -\frac{1}{2M^2_{pl}}{\cal G}_{IJ}\dot{\phi}^I\dot{\phi}^J,\label{conteq}
\end{eqnarray}
where $H = \dot a/a$ and an overdot denotes taking the derivative with
respect to time.  The equations of motion for the homogeneous fields
$\phi^I$ are given as
\begin{equation}
 \mathcal D_t\dot{\phi}^I + 3H\dot{\phi^I} 
+ {\cal G}^{IJ}V_{,J} = 0,
  \label{fldeq}
\end{equation}
where $V_{,J}= \partial V/\partial\phi^J$ and we have introduced the
covariant time derivative $\mathcal D_t$ that acts as $\mathcal D_t X^I
= \dot X^I + \Gamma^I_{\,JK}\dot\phi^JX^K$, with the Christoffel symbols
$\Gamma^I_{\;JK}$ being constructed from ${\cal G}_{IJ}$ and its
derivatives (see Appendix for details).  In our case, the equations of
motion for $\phi$ and $\chi$ are given as
\begin{gather}
 \ddot{\phi} + 3H\dot{\phi} + \alpha e^{-2\alpha\phi}\dot{\chi}^2 +
 V_{,\phi} = 0,\\
 \ddot{\chi} + 3H\dot{\chi} - 2\alpha \dot{\phi}\dot{\chi} + e^{2\alpha\phi}V_{,\chi} = 0.
\end{gather}

In the context of inflation, it is useful to define the slow-roll
parameters $\epsilon = -\dot H/H^2$ and $\eta = \dot \epsilon/(\epsilon
H)$.  In order to obtain quasi-exponential inflation we require
$\epsilon\ll 1$, and the condition $\eta\ll 1$ ensures that inflation
lasts for long enough.\footnote{Strictly speaking $\eta$ can be
negative. So we take slow-roll to mean that $|\eta|\ll 1$.}  The amount
of inflation is parameterised in terms of the $e$-folding number $N$
defined as
\begin{align}
 N(t,t_\ast) = \int^t_{t_\ast}H(t)dt = \ln\left(\frac{a(t)}{a(t_\ast)}\right), 
\end{align}
where $t_\ast$ is the initial time, and observational constraints
dictate that $N\gtrsim 60$.  
In terms of the scalar fields, we have  
\begin{align}
 \epsilon = \frac{1}{2M_{pl}^2}\frac{\mathcal
 G_{IJ}\dot\phi^I\dot\phi^J}{H^2}\qquad\mbox{and}\qquad \eta = 2\epsilon
 + 2\frac{\mathcal G_{IJ}\dot\phi^I\mathcal D_t\dot\phi^J}{H\mathcal G_{KL}\dot\phi^K\dot\phi^L}.
\end{align}   
The slow-roll condition $\epsilon\ll 1$ thus implies that 
\begin{align}
 H^2 \simeq \frac{V(\phi^I)}{3M_{pl}^2}.\label{srfrdmn}
\end{align}
Given that $\epsilon\ll 1$, the condition $\eta\ll 1$ implies that
\begin{align}
 \frac{\mathcal G_{IJ}\dot\phi^I\mathcal D_t\dot\phi^J}{H\mathcal
 G_{KL}\dot\phi^K\dot\phi^L}\ll 1.\label{cond1}
\end{align}
In the single-field case, where we can always redefine the field such
that $\mathcal G_{\phi\phi} = 1$, this reduces to $\ddot\phi \ll
H\dot\phi$, which allows us to neglect the acceleration term in the
equation of motion for $\phi$.  In the multi-field case with a curved
field space, however, the situation is not so simple, as the above
condition only constrains the component of $\mathcal D_t\dot\phi^I$
along the background trajectory.  Nevertheless, we assume that the
magnitude of the acceleration vector $\mathcal D_t\dot\phi^I$ is much
smaller than the magnitude of the velocity vector $H \dot\phi^I$, namely
$(\mathcal G_{IJ}\mathcal D_t\dot\phi^I\mathcal D_t\dot\phi^J)^{1/2}\ll
H(\mathcal G_{IJ}\dot\phi^I\dot\phi^J)^{1/2}$.  By the Cauchy-Schwarz
inequality, this will guarantee that condition \eqref{cond1} is
satisfied.  If we further assume that the field basis is such that
$|\mathcal D_t\dot\phi^I|\ll |H\dot\phi^I|$ for all $I$, where here by
$|X^I|$ we mean the magnitude of the $I$th component of $X^I$, then the
equations of motion \eqref{fldeq} reduce to
\begin{equation}
  3H\dot{\phi^I} \simeq - {\cal G}^{IJ}V_{,J},
  \label{srfldeq}
\end{equation}
meaning that we are in an attractor regime where the field velocities
are given as functions of the field positions.  Using the slow-roll
equations \eqref{srfrdmn} and \eqref{srfldeq} we can then derive
consistency conditions for the potential and its derivatives.  Namely,
we find
\begin{align}
 \epsilon \simeq \epsilon_V = \frac{M_{pl}^2}{2}\frac{\mathcal
 G^{IJ}V_{,I}V_{,J}}{V^2}\ll1,\qquad \eta \simeq
 \eta_V = 4\epsilon_V -\frac{M_{pl}^4}{\epsilon_V}\frac{\mathcal D_KV_{,J}\mathcal
 G^{KL}V_{,L}\mathcal G^{JM}V_{,M}}{V^3}\ll 1, 
\end{align}
where $\mathcal D_KV_{,J}= V_{,JK} - \Gamma^L_{\,JK}V_{,L}$ is the
covariant derivative of $V_{,J}$.  The first condition $\epsilon_V\ll 1$
thus puts a constraint on the first derivatives of $V$, while the
condition $\eta_V\ll 1$ constrains the second derivatives of $V$.  In
particular, assuming $\epsilon_V\ll 1$, the condition $\eta_V\ll 1$ will
be satisfied if we assume that all the eigenvalues of the field-space
tensor $\eta^I_{\,J}$ are small, where $\eta^I_{\,J}$ is defined as
\begin{align}
 \eta^I_{\,J}\equiv M_{pl}^2 \frac{\mathcal G^{IK}\mathcal D_JV_{,K}}{V}.
\end{align}
Provided the number of fields is not too large, the Eigenvalues of
$\eta^I_{\,J}$ will in turn be small if $\eta^I_{\,J}\ll 1$ for all $I$
and $J$.  Note that the quantity $\eta^I_{\,J}$ corresponds to the
covariant Hessian of the potential divided by $V/M_{pl}^2\simeq 3H^2$,
and the covariant Hessian of the potential contributes to the effective
mass matrix of field fluctuations about the background trajectory, see
e.g.~\cite{Sasaki:1995aw}.  As such, the condition $\eta^I_{\,J}\ll 1$
will constitute part of the sufficient condition for the effective mass
of field fluctuations to be small compared to the Hubble scale.

%
%
\section{$\zeta$ and its correlation functions using the $\delta N$ formalism}
\label{PSBS}

Having introduced the model and background equations in the previous
section, we now move on to consider perturbations.  In particular, we
are interested in determining the so-called curvature perturbation on
constant density slices, $\zeta$, as it is this quantity that can be
related to the temperature fluctuations observed in the CMB.  More
precisely, we are interested in the correlation functions of $\zeta$, as
it is only the statistical properties of the CMB temperature
fluctuations that we are able to make predictions for.

In general, following the notation of \cite{Sugiyama:2012tj}, the full
perturbed metric can be written in the form
\begin{equation}
 ds^2 = -\alpha^2dt^2 + a^2(t)e^{2\psi}\gamma_{ij}(dx^i +
  \beta^idt)(dx^j + \beta^j dt),\label{pertmet}
\end{equation}
where $\gamma_{ij}$ has unit determinant and can be written in terms of
the traceless tensor $h_{ij}$ as $\gamma_{ij} = (e^h)_{ij}$.  $h_{ij}$
can itself be decomposed as
\begin{align}
 h_{ij} = \partial_iC_j + \partial_jC_i
 -\frac{2}{3}\delta_{ij}\partial_k C_k + h_{ij}^{(T)},
\end{align}
where $C_i$ contains both scalar and vector perturbations and
$h_{ij}^{(T)}$, satisfying $\partial_i h_{ij}^{(T)} = 0$, corresponds to
tensor perturbations.  In addition to metric perturbations, we also have
the perturbed scalar fields $\hat\phi^I(t,\bm x) = \phi^I(t) +
\delta\phi^I(t,\bm x)$.  The density associated with their
energy-momentum tensor is similarly decomposed as $\hat\rho(t,\bm x) =
\rho(t) + \delta\rho(t,\bm x)$.  Following \cite{Sugiyama:2012tj}, we
make use of the flat threading, in which $C_i = 0$, and the curvature
perturbation on constant density slices, $\zeta$, then corresponds to
$\psi$ evaluated in the gauge where $\delta\rho = C_i = 0$, namely
$\zeta = \psi|_{\delta\rho = C_i = 0}$.

\subsection*{Separate Universe approach and the $\delta N$ expansion}

In order to determine $\zeta$ and its correlation functions we make use
of the separate Universe approach and the $\delta N$ formalism
\cite{Sasaki:1995aw,Sasaki:1998ug,Wands:2000dp,Lyth:2004gb,Lyth:2005fi,Sugiyama:2012tj}.
The separate Universe approach corresponds to the leading order
approximation in a gradient expansion.  One first assumes that the
characteristic length scale of spatial variations, $L$, is longer than
the Hubble scale, namely $\xi = 1/(HL)\ll 1$.  Associating a factor of
$\xi$ with spatial gradients appearing in the field equations, one can
then perform an expansion in the parameter $\xi$.  Neglecting terms of
order $\xi^2$ and higher, one finds that the field equations take on
exactly the same form as the background equations.  In other words,
separate super-Hubble sised patches are found to evolve as separate
background Universes, differing only in their initial conditions.  If we
are interested in a comoving scale with wavenumber $k$, during inflation
the parameter $\xi= k/(aH)$ will be decreasing exponentially with time.
As such, the separate Universe approach will become applicable after the
Horizon-crossing time, which is defined as the time at which $k = aH$.
 
Making use of the flat gauge, corresponding to $\psi= C_i =0$, the
validity of the separate Universe approach in the case of multiple
scalar fields has been confirmed explicitly to all orders in
perturbation theory by Sugiyama {\it et al.} \cite{Sugiyama:2012tj}.
$\beta^i$ and $\dot h_{ij}^{(T)}$ were shown to decay away on
super-horizon scales, such that the field equations indeed take on
exactly the same form as the background equations, namely
\begin{gather}\label{pertFrdmn}
 \hat H^2 =
 \frac{1}{3M_{pl}^2}\left[\frac{1}{2}\partial_\tau\hat\phi^I\partial_\tau\hat\phi^J
 + V\left(\hat\phi^I\right)\right],\\
\mathcal D_\tau\partial_\tau\hat\phi^I + 3\hat H\partial_\tau \hat\phi^I
 +\mathcal G^{IJ}V_{,J}(\hat\phi^K) = 0 \label{pertEoM}
\end{gather}
where $\hat H(t,\bm x) = H(t)/\alpha(t,\bm x)$ is the local Hubble
expansion and $\partial_\tau = \partial/\partial\tau$, with $d\tau =
\alpha(t,\bm x)dt$.  A result that proves very useful is that the local
$e$-folding number is found to be unperturbed \cite{Sasaki:1995aw,Lyth:2004gb}, as
\begin{align}
 \hat N = \int\hat H d\tau = \int H dt.
\end{align}
This can also be understood if, associated with the perturbed metric in
eq.~\eqref{pertmet}, we define the effective scale factor $\hat a(t,\bm
x) = a(t)e^{\psi(t,\bm x)}$.  The local $e$-folding number is then given
as
\begin{align}
 \hat N = \ln\left(\frac{\hat a}{\hat a_\ast}\right) = \psi - \psi_\ast + N,\label{localefold}
\end{align}
and in the flat slicing, i.e. $\psi = \psi_\ast = 0$, this reduces to
the background $e$-folding number.  The $e$-folding number is thus a useful time
parameter in the flat gauge, and given that the field equations
\eqref{pertFrdmn} and \eqref{pertEoM} take on the same form as the
background equations, we are able to write
\begin{align}
 \hat\phi^I(N,\bm x) = \phi^I(N,\phi_\ast^J(\bm x)),\label{susol}
\end{align}
where $\phi^I(N,\phi_\ast^J(\bm x))$ is a solution of the background
equations of motion with the spatially dependent initial conditions
$\phi^I(t_\ast) = \phi^I_\ast(\bm x)$.\footnote{In principle we also
need to specify the initial field velocities, but we will assume that
the slow-roll approximation is valid around the time of horizon
crossing, such that field velocities are given in terms of the field
values as in eq.~\eqref{srfldeq}.}  In other words, the value of
$\hat\phi^I$ at a given location $\bm x$ is found simply by solving the
background equations of motion with the appropriate initial conditions
for that location.

Having outlined the separate Universe approach, we now wish to determine
$\zeta$, and for this we use the $\delta N$ formalism
\cite{Sasaki:1995aw,Sasaki:1998ug,Lyth:2004gb,Sugiyama:2012tj}. The
basic idea of the $\delta N$ formalism is that $\zeta$ on some final
uniform density slice can be given in terms of the spatial fluctuations
of the $e$-folding number between an initial spatially flat slice and
the final uniform density slice.  This can be understood if we look at
the expression for the local $e$-folding number given in
\eqref{localefold}.  Taking the initial slice to be flat and the final
slice to be a constant density one, corresponding to $\psi_\ast = 0$ and
$\psi = \zeta$, we find $\delta N = \hat N - N = \zeta$.  Note that it
does not matter exactly when we take the initial flat slice, as it is
only important that $\psi_\ast = 0$.  The only restriction is that
$t_\ast$ must be after the time at which the scales under consideration
have left the horizon.

The next step in the $\delta N$ formalism is to show that $\delta N$ can
be expanded in terms of the field perturbations on the initial flat
slice, $\delta\phi^I_\ast(\bm x)$.  To see this, recall that in the
context of the separate Universe approach the field equations take on
exactly the same form as the background equations.  As mentioned above,
this means that the solutions for $\hat\phi^I(N,\bm x)$ in the flat
gauge are as given in eq.~\eqref{susol}, i.e. they are solutions to the
background equations but with the initial conditions varying from place
to place.  Similarly, it also means that the energy density on flat
slices can be expressed as
\begin{align}
 \hat\rho(N,\bm x) = \rho(N,\phi^I_\ast(\bm x)), 
\label{rhophirel}
\end{align}
where $\rho(N,\phi^I_\ast(\bm x))$ is the density as determined by
solving the background field equations with the spatially inhomogeneous
initial conditions $\phi^I(t_\ast) =\phi^I_\ast(\bm x)$.  In general
$\hat\rho(N,\bm x)$ is not spatially homogeneous, and if we assume that
the initial conditions at position $\bm x$ can be expanded about the
initial conditions of the fiducial background trajectory as
$\phi^I_\ast(\bm x) = \phi^I_\ast + \delta\phi^I_\ast(\bm x)$, this
leads to an expansion of the form
\begin{align}
\hat\rho(N,\bm x) =\rho(N,\phi^I_\ast) + \rho_{,I}(N,\phi^J_\ast)\delta\phi^I_\ast(\bm x) +
 \frac{1}{2}\rho_{,IJ}(N,\phi^K_\ast)\delta\phi^I_\ast(\bm x) \delta\phi^J_\ast(\bm
 x) + ...\, ,  \label{rhoexp}
\end{align}
where $\rho(N,\phi^I_\ast)$ is the density of the fiducial background
trajectory, $\rho_{,I}(N,\phi^J_\ast) =
\partial\rho(N,\phi^J_\ast)/\partial\phi^I_\ast$ and similarly for
$\rho_{,IJ}(N,\phi^K_\ast)$.  At each location $\bm x$, we can then
consider the shift along the local trajectory, $\delta N$, that is
required to reach a constant density slice.  In other words, at each
location $\bm x$ we find the shift in $N$ such that $\hat\rho(N+\delta
N,\bm x) = \rho(N,\phi^I_\ast)$.  Given the form of the expansion for
$\hat\rho(N,\bm x)$ in eq.~\eqref{rhoexp}, solving $\hat\rho(N+\delta
N,\bm x) = \rho(N,\phi^I_\ast)$ gives rise to an expansion of the form
\begin{equation}
 \zeta(N,\bm x) = \delta N(N,\bm x) = N_{,I}(N,\phi^J_\ast)\delta\phi^I_\ast(\bm x) + \frac{1}{2} N_{,IJ}(\phi^K_\ast)\delta\phi^I_\ast(\bm x)\delta\phi^J_\ast(\bm x) + \cdots,
  \label{z_exp}
\end{equation}
which is the famous $\delta N$ expansion.  As mentioned above,
in principle the initial flat slice can be taken to be at any time after
the scales under consideration have left the horizon, but in practice it
is useful to choose it to coincide with the horizon crossing time, as
expressions for the quantities $\delta\phi^I_\ast(\bm x)$ and their
correlations at this time are known \cite{Sasaki:1995aw,Nakamura:1996da}.

While the above form for the expansion of $\zeta$ is perfectly
acceptable, in the case of a curved field space the field perturbations
$\delta\phi^I_\ast(\bm x)= \phi^I_\ast(\bm x) - \phi^I_\ast$, which
correspond to coordinate displacements, do not transform covariantly.
In order to obtain an explicitly covariant expression for $\zeta$ we
follow the discussion in \cite{Gong:2011uw}, see also \cite{Elliston:2012ab,Kaiser:2012ak}.  For
sufficiently small $\delta\phi^I_\ast(\bm x)$, the two points in field
space $\phi^I_\ast(\bm x)$ and $\phi^I_\ast$ are connected by a unique
geodesic that we take to be parameterised by $\lambda$.  Normalising
$\lambda$ such that $\phi^I(\lambda=0)=\phi^I_\ast$, and
$\phi^I(\lambda=1)=\phi^I_\ast(\bm x)$, we can obtain a Taylor series
expansion for $\delta\phi^I = \phi^I(\lambda = 1) - \phi^I(\lambda = 0)$
as
\begin{align}
 \delta\phi^I = \left.\frac{d\phi^I}{d\lambda}\right|_{\lambda = 0} +
 \frac{1}{2}\left.\frac{d^2\phi^I}{d\lambda^2}\right|_{\lambda = 0} + \cdots.
\end{align}
On the other hand, the geodesic satisfies
\begin{align}
 \mathcal D_\lambda \frac{d\phi^I}{d\lambda} \equiv
 \frac{d^2\phi^I}{d\lambda^2} +
 \Gamma^I_{\,JK}\frac{d\phi^J}{d\lambda}\frac{d\phi^K}{d\lambda} = 0.
\end{align}
As such, introducing $\mathcal Q^I = d\phi^I/d\lambda|_{\lambda = 0}$, which
resides in the tangent space at $\phi^I(\lambda = 0)$ and thus
transforms covariantly,  we can express $\delta\phi^I$ in terms of $\mathcal
Q^I$ as 

\begin{equation}
 \delta \phi^I = \mathcal Q^I - \frac{1}{2!} \Gamma^I_{\,JK}\mathcal Q^J \mathcal Q^K 
  + \cdots.
\end{equation}
Inserting this relation into \eqref{z_exp} we obtain
\begin{equation}
 \zeta(N,\bm x) =  N_{,I}(N,\phi_\ast^J) \mathcal Q^I_\ast(\bm x) +
  \frac{1}{2} {\mathcal D}_I\mathcal D_J N(N,\phi_\ast^K)\mathcal Q^I_\ast(\bm x)\mathcal Q^J_\ast(\bm x) + \cdots,\label{covzeta}
\end{equation}
which is now explicitly covariant.

\subsection*{The power spectrum and bispectrum of $\zeta$}

Having obtained an expansion for $\zeta$ in terms of the covariantised
field perturbations on a flat slice at the horizon crossing time, we now
turn to the correlation functions of $\zeta$.  Working in Fourier
space, the two-point correlation function of $\zeta$ is parameterised as
\begin{equation}
 \la\zeta(\bkone)\zeta(\bktwo)\ra = (2\pi)^3\delta^3(\bkone + \bktwo)P_{\zeta}(k_1)
  = (2\pi)^3\delta(\bkone + \bktwo)\frac{2\pi^2}{k_1^3}{\cal P}_{\zeta}(k_1),
  \label{ps_def} 
\end{equation}
and the three-point correlation function is similarly parameterised as
\begin{equation}
 \begin{aligned}
  \la\zeta(\bkone)\zeta(\bktwo)\zeta(\bkthree)\ra
  &= (2\pi)^3\delta^3(\bkone + \bktwo + \bkthree)B_{\zeta}(k_1,k_2,k_3).
  \label{B_def}
 \end{aligned} 
\end{equation}
$P_\zeta(k)$ and $\mathcal P_\zeta(k)$ are the power spectrum and
reduced power spectrum, respectively, while $B_\zeta(k_1,k_2,k_3)$ is
the bispectrum.  In both \eqref{ps_def} and \eqref{B_def} the delta
functions are a consequence of assuming statistical homogeneity, and the
fact that $P_\zeta$, $\mathcal P_\zeta$ and $B_\zeta$ depend only on the
magnitudes of $\bm k_i$ is a consequence of assuming statistical
isotropy.  In relation to the three-point function, a useful parameter
introduced to quantify the level of non-Gaussianity is $f_{NL}$, which
is defined as
\begin{equation}
 f_{\rm NL} =\frac{5}{6} \frac{B_{\zeta}(k_1,k_2,k_3)}{ P_{\zeta}(k_1)P_{\zeta}(k_2) + {\rm c.p.} },
  \label{fNL_def}
\end{equation}
where ${\rm c.p.}$ denotes cyclic permutations of $k_1$, $k_2$ and
$k_3$.  Given the expansion for $\zeta$ in eq.~\eqref{covzeta}, we see
that the correlation functions of $\zeta$ can be expressed in terms of
the correlation functions of the covariantised field perturbations on
the initial flat slice, $\mathcal Q^I$.  In particular, we have
\begin{align}
\langle\zeta(\bkone)\zeta(\bktwo)\rangle &= N_{,I}N_{,J}\la\mathcal Q^I_\ast(\bkone)\mathcal Q^J_\ast(\bktwo)\ra,\label{2ptfct}\\
\la\zeta(\bkone)\zeta(\bktwo)\zeta(\bkthree)\ra
  &=N_{,I}N_{,J}N_{,K} \la {\cal Q}^I_\ast(\bkone){\cal Q}^J_\ast(\bktwo){\cal Q}^K_\ast(\bkthree) \ra\label{3ptfct} \\\nonumber
  &\quad + N_{,I} N_{,J} {\cal D}_K{\cal D}_L N
  \int\frac{d^3 {\bf q}}{(2\pi)^3}\la {\cal Q}^K_\ast(\bkone-{\mathbf q}){\cal Q}^I_\ast(\bktwo) \ra
  \la{\cal Q}^L_\ast({\mathbf q}) {\cal Q}^J_\ast(\bkthree) \ra   +   {\rm c.p}\,,
\end{align}
where for brevity we drop the arguments of $N_{,I}$ and $\mathcal
D_J\mathcal D_IN$.  The contribution to the three-point correlation
function of $\zeta$ coming from the first term involving the three-point
correlation functions of $\mathcal Q^I$ is known to be unobservably
small \cite{Seery:2005gb,Lyth:2005qj}, so in proceeding we choose to
neglect it.  As such, the only quantities required are the two-point
correlation functions of $\mathcal Q^I$.  At linear order in
perturbations we have $\mathcal Q^I = \delta\phi^I$, and the two-point
correlation functions of $\delta\phi^I$ in the case of a curved field
space have been calculated in \cite{Sasaki:1995aw,Nakamura:1996da}.  The
result at lowest order in slow-roll is
\begin{align}
 \la\mathcal Q^I_\ast(\bkone)\mathcal Q^J_\ast(\bktwo)\ra =
 (2\pi)^3\delta^3(\bkone + \bktwo)\frac{2\pi^2}{k_1^3} \left(\frac{H_\ast}{2 \pi}\right)^2\mathcal G^{IJ}_\ast,
\end{align}
where recall that an asterisk now denotes that a quantity should be
evaluated at the time of horizon crossing, namely $k_1 = a_\ast H_\ast$.
Substituting this result into eqs.~\eqref{2ptfct} and \eqref{3ptfct},
expressions for $\mathcal P_\zeta(k_1)$ and $f_{NL}$ are obtained as
\begin{align}
 \mathcal P_\zeta(k) = \left(\frac{H_\ast}{2\pi}\right)^2\mathcal
 G^{IJ}_\ast N_{,I}N_{,J},\label{powcov}\\
f_{NL} =  \frac{5}{6} \frac{ N^{,I} N^{,J} {\cal D}_I {\cal D}_J N}{ \left(N^{,K} N_{,K} \right)^2},\label{fNLcov}
\end{align}
where the raised indices in the second expression are raised with
$\mathcal G^{IJ}_\ast$.

In addition to the above two observables, we also consider the tilt of
the power spectrum, $n_s$, and the tensor-to-scalar ratio, $r$.  The
tilt of the power spectrum is defined through the relation
\begin{equation}
 {\cal P}_{\zeta}(k) = A_s \left(\frac{k}{k_p}\right)^{n_s-1},
\end{equation}
where $k_p$ is some pivot scale and $A_s$ gives the magnitude of
$\mathcal P_\zeta$ at the pivot scale.  The scale dependence of
$\mathcal P_\zeta$ as given in eq.~\eqref{powcov} appears through its
dependence on quantities evaluated at the horizon crossing time of the
comoving scale $k$.  We do not present a detailed derivation here, but
the final result is given as \cite{Sasaki:1995aw}
\begin{equation}
  n_s = 1 -2\epsilon_\ast
    - 2\frac{1+N_{,I}\left(\frac{1}{3}R^{IJKL}\frac{V_{,J}V_{,K}}{V^2} -
		      \frac{{\cal D}^I \mathcal D^LV}{V}\right)_\ast N_{,L}}{N^{,M}N_{,M}},\label{tilt}
\end{equation}
where, $R^{IJKL}$ is the curvature tensor constructed from ${\cal
G}_{IJ}$.  Note that $A_s$ is found simply by taking
$k=k_p$ in eq.~\eqref{powcov}.  

Finally, the tensor-to-scalar ratio is defined as the ratio between the
power spectra of tensor and scalar perturbations.  In particular, if we
parameterise the power spectrum of tensor perturbations as
\begin{align}
 \mathcal P_T(k) = A_T\left(\frac{k}{k_p}\right)^{n_T},
\end{align}
then we have $r = A_T/A_s$.  It can be shown that $A_T =
8(H_\ast/(2\pi))^2/M_{pl}^2$, see e.g. \cite{Bassett:2005xm}, such that we obtain 
\begin{equation}
 r = \frac{8}{M_{pl}^2N^{,I}N_{,I}}. \label{ratio} 
\end{equation}

\subsection*{Observational bounds}

In general, as indicated in eq.~\eqref{covzeta}, $\zeta$ will depend on
time.  As such, when trying to compare the predictions of a particular
model with observational constraints it is important that we choose an
appropriate time at which to evaluate $\zeta$.  In the case of
single-field inflation it is known that even for a very general class of
models $\zeta$ is conserved on super-horizon scales, see
e.g. \cite{Naruko:2011zk,Gao:2011mz}, which means that the appropriate
time to evaluate $\zeta$ is shortly after the scales under consideration
left the horizon.  In the multi-field case, however, $\zeta$ can
continue to evolve on super-horizon scales, and so it becomes necessary
to follow the evolution of $\zeta$ up until a so-called adiabatic limit
is reached and $\zeta$ becomes conserved.\footnote{It is also possible
that a so-called adiabatic limit is not reached, but non-adiabatic
contributions to the CMB temperature fluctuations are now tightly
constrained \cite{Ade:2015lrj}.}  The non-conservation of $\zeta$ during
inflation results from it being sourced by so-called isocurvature
perturbations when the background trajectory deviates from a geodesic of
the field space, see e.g.~\cite{Gordon:2000hv,Langlois:2008mn}.
Isocurvature perturbations are field perturbations orthogonal to the
background trajectory, and an adiabatic limit corresponds to the
situation where all isocurvature perturbations have decayed away,
leaving only perturbations along the trajectory.  With no isocurvature
modes to source $\zeta$ it becomes conserved, and the fact that only
perturbations along the trajectory --- adiabatic perturbations ---
remain means that an adiabatic limit corresponds to an effectively
single-field limit.  If isocurvature modes have not decayed away by the
end of inflation, then one must continue to following the evolution of
$\zeta$ through (p)reheating.  An analysis of (p)reheating in the model
under consideration, however, is beyond the scope of this paper, and
will very much depend on how the fields $\phi$ and $\chi$ in
eq.~\eqref{Eaction} are coupled to other forms of matter.\footnote{One
choice of matter and coupling is considered in
\cite{vandeBruck:2016leo}.}  As such, in the following we will focus on
the evolution of $\zeta$ during inflation and its properties at the end
of inflation.  In cases where an adiabatic limit is reached before the
end of inflation we are justified in comparing our results with
observational constraints, and we will point out when this is not the
case and evolution of $\zeta$ through (p)reheating may be important.

Closely related to the issue of whether or not an adiabatic limit is
reached is the issue of frame-dependence.  Most of our discussion up to
now has been centred on the Einstein-frame representation of the model
given in eq.~\eqref{Eaction}, but we could equally have chosen to
perform our analysis in the original Jordan-frame representation.  At
the classical level the two representations simply correspond to a
re-labelling of the metric, and if we perform calculations consistently
then predictions for observable quantities should be independent of the
choice of frame.  The physical picture in the two frames, however, may be
very different, see e.g.~\cite{Deruelle:2010ht}.  Indeed, in the context
of inflation, at the level of the background the definition of inflation
is not even a frame-independent notion, and it is possible to have a
situation where the FLRW metric in one frame is inflating while that in
the other is not, see e.g.~\cite{Domenech:2015qoa}.  At the level of
perturbations, however, the situation is fortunately somewhat simpler.
Tensor perturbations are left unchanged by a conformal transformation,
and $\zeta$ is also frame-independent in the single-field case or in the
case that an adiabatic limit has been reached
\cite{Makino:1991sg,Chiba:2008rp,Gong:2011qe}.  Only in the case that
isocurvature perturbations remain does one have to be a little more
careful, as $\zeta$ is not necessarily frame-independent in this case
\cite{White:2012ya,White:2013ufa,Chiba:2013mha}.  Note that this does
not imply an inequivalence between the two frames, but simply indicates
that the quantity $\zeta$ itself does not yet directly correspond to an
observable quantity.  As discussed above, independent of the frame issue
it becomes necessary to follow the evolution of $\zeta$ through
(p)reheating in the case that an adiabatic limit is not reached by the
end of inflation, and we will not address this issue here.  In this
sense, the potential frame-dependence of $\zeta$ will only affect the
cases for which we are already not justified in comparing our results
with observations, so we postpone addressing the issue in any more detail.
Note that for our particular model the Jordan and Einstein frames will
coincide once the field $\phi$ relaxes to $\phi = 0$.

With the above remarks in mind, the observational constraints on $A_s$,
$n_s$, $r$ and $f_{NL}$ with which we compare our results are those
presented by the {\it Planck} collaboration~\cite{Ade:2015lrj}.  Taking
a pivot scale of $k_p=0.05\,{\rm Mpc}^{-1}$ they find
\begin{gather}
 A_s = (2.21 \pm 0.07) \times 10^{-9} \ \ ({\rm 68\% \; C.L.}),\\
n_s = 0.968 \pm 0.006 \ \ ({\rm 68\% \; C.L.}), \\   
  r < 0.11 \ \ ({\rm 95\% \; C.L.}),\\
f_{\rm NL} = 0.8 \pm 5.0 \ \ ({\rm 68\% \; C.L.}).
\end{gather}

%
%
\section{Numerical analysis and results}
\label{Num}

As can be seen from the expressions given in
eqs.~\eqref{powcov},~\eqref{fNLcov},~\eqref{tilt} and \eqref{ratio}, the
observables $\mathcal P_\zeta$, $n_s$, $r$ and $f_{NL}$ for a given
inflationary trajectory can be determined with knowledge of the
background dynamics alone, which is one of the very appealing aspects of
the $\delta N$ formalism.  In particular, we require the background
quantities $H$, $\epsilon$, $\mathcal G_{IJ}$ and $V$ evaluated at the
horizon-crossing time of the comoving scale under consideration, as well
as the derivatives of the $e$-folding number up to a constant density
surface with respect to the field values at the horizon crossing time,
$N_{,I}$ and $\mathcal D_J \mathcal D_IN$.  For a restricted class of
potentials and field-space metrics it is possible to determine the
derivatives of $N$ analytically if the slow-roll equations of motion
\eqref{srfldeq} are assumed to hold throughout inflation, see e.g.
\cite{Vernizzi:2006ve,Choi:2007su,Sasaki:2008uc,Meyers:2010rg,Elliston:2011dr},
but in general one has to resort to numerical calculations.  The model
we are considering contains both a non-trivial field-space metric and
interaction terms in the potential, meaning that the derivatives of $N$
cannot be determined analytically.  We thus have to take a numerical
approach, the method of which we now briefly explain.

Our code is based on the finite difference method.  We first consider a
background trajectory with the initial conditions
$(\phi_\ast,\chi_\ast)$, and assume that the scale under consideration
left the horizon as the trajectory passed through this point.  Evolving
along the trajectory, at any later time of interest $t$, we can
determine the number of $e$-foldings since the horizon-crossing time,
$N(t,\phi_\ast,\chi_\ast)$, and the density at that time,
$\rho(t,\phi_\ast,\chi_\ast)$.  Next we consider another trajectory with
displaced initial conditions, e.g. $(\phi_\ast + \Delta
\phi,\chi_\ast)$.  Evolving along this trajectory we determine the time
$\tilde t = t + \delta t$ at which the density of the displaced
trajectory coincides with $\rho(t,\phi_\ast,\chi_\ast)$, namely
$\rho(\tilde t,\phi_\ast + \Delta\phi,\chi_\ast) =
\rho(t,\phi_\ast,\chi_\ast)$.  We then determine the number of
$e$-foldings that have elapsed on the perturbed trajectory from the
initial time up to the time $\tilde t$, $N(\tilde t,\phi_\ast +
\Delta\phi,\chi_\ast)$.  This is the number of $e$-foldings up to the
constant density surface, and the derivative of $N$ with respect to
$\phi_\ast$ is then given as
\begin{equation}
 N_{,\phi_\ast} = \frac{N(\tilde t,\phi_*+\Delta\phi,\chi_*)-N(t,\phi_*,\chi_*)}{\Delta\phi}.
\end{equation}
The same procedure applies for determining $N_{,\chi_\ast}$, and can be
extended to calculate the second-order derivatives
$N_{,\phi_\ast\phi_\ast}$, $N_{,\chi_\ast\chi_\ast}$ and
$N_{,\phi_\ast\chi_\ast} = N_{,\chi_\ast\phi_\ast}$.  In all our
calculations we assume that the slow-roll field equations
\eqref{srfrdmn} and \eqref{srfldeq} are a good approximation at the time
of horizon crossing.  The initial field velocities are thus determined
through eq.~\eqref{srfldeq} and do not need to be specified
independently.  Nevertheless, we do solve the full equations of motion
eq.~\eqref{fldeq} when calculating the derivative of $N$.  This allows
for the possibility that the slow-roll approximation breaks down later
on during the super-horizon evolution.  We have worked with 32-digit
precision.

Note that the time $t$ in the above discussion can be any time after the
horizon crossing time, so by varying $t$ we can determine the evolution
of $\zeta$.  If we are interested in determining $\zeta$ at the end of
inflation, then we take $t$ to be the time at which $\epsilon \simeq 1$.
As discussed at the end of the previous section, if an adiabatic limit
has not been reached by the end of inflation then it is necessary to
follow the evolution of $\zeta$ through (p)reheating and until an adiabatic
limit is reached and $\zeta$ becomes conserved.  However, the evolution
of $\zeta$ through (p)reheating is beyond the scope of this paper, and we
restrict our attention to the evolution of $\zeta$ up until the end of
inflation.

In proceeding, rather than working with the parameters $m_\phi$ and
$m_\chi$, we instead introduce the mass ratio defined as
\begin{align}
 R_{\rm mass} \equiv \frac{m_\chi}{m_\phi}.
\end{align}
This allows an overall $m_\phi^2$ to be factored out of the potential,
namely
\begin{align}
  V(\phi,\chi) = m_\phi^2\mathcal V(\phi,\chi),\qquad \mathcal
 V(\phi,\chi) = \frac{3}{4} M_{pl}^2(1-e^{-2\alpha \phi})^2 +
 \frac{1}{2} R_{\rm mass}^2 e^{-4\alpha \phi} \chi^{2}.
  \label{scaledpot}
\end{align}
If we then introduce the re-scaled time parameter $\tilde\tau = m_\phi t$, we
find that the background field equations reduce to
\begin{gather}
 \mathcal D_{\tilde\tau}\phi^I_{\tilde\tau} + 3\mathcal H\phi^I_{\tilde \tau} 
+ {\cal G}^{IJ}\mathcal V_{,J} = 0,\\ \mathcal H^2 = \frac{1}{3M^2_{pl}} \bigg[ \frac{1}{2}{\cal
G}_{IJ}\phi^I_{\tilde\tau}\phi^J_{\tilde\tau} + \mathcal V(\phi^I) \bigg],
  \label{scaledfldeq}
\end{gather}
where a subscript $\tilde\tau$ denotes taking the derivative with
respect to $\tilde\tau$, e.g. $\phi^I_{\tilde\tau} =
d\phi^I/d\tilde\tau$, $\mathcal H = a_{\tilde\tau}/a$ and $\mathcal
D_{\tilde\tau} X^I = X^I_{\tilde\tau} +
\Gamma^I_{\,JK}\phi^J_{\tilde\tau} X^K$.  As such, we see that the mass
$m_\phi$ drops out of the field equations.  In particular, this means
that the solution for $\mathcal H$ as a function of $\tilde\tau$ will be
independent of $m_\phi$.  If we then consider the definition of the
$e$-folding number, we have
\begin{align}
 N = \int^t_{t_\ast} H dt = \int^{\tilde\tau}_{\tilde\tau_\ast} \mathcal H d\tilde\tau,
\end{align}
from which we conclude that the $e$-folding number is independent of the
overall mass scale $m_\phi$.  This in turn means that the derivatives of
$N$, required in calculating $\mathcal P_\zeta$, $n_s$, $r$ and
$f_{NL}$, will also be independent of $m_\phi$.  Given the expressions
for $n_s$, $r$ and $f_{NL}$, we thus find that they are all independent
of $m_\phi$.  The only observable that depends on $m_\phi$ is $\mathcal
P_\zeta$, as this depends on the overall normalization of
$H_\ast^2$.\footnote{Another way to see the independence of $m_\phi$ is
to write the field equations of motion directly in terms of the time
parameter $N$.  On doing so, the potential only appears in the
combination $V_{,I}/V$, meaning that the overall factor of $m_\phi^2$
drops out.}  Explicitly, we have
\begin{align}
 \mathcal P_\zeta(k) = m_\phi^2\left(\frac{\mathcal H_\ast}{2\pi}\right)^2\mathcal
 G^{IJ}_\ast N_{,I}N_{,J},
\end{align}  
which means that we are able to determine the quantity $\mathcal
P_\zeta/m_\phi^2$ without knowing $m_\phi$.

Using this new parameterization, the free parameters of the theory
(assuming slow-roll at horizon crossing) are $m_\phi$, $R_{\rm mass}$,
$\phi_\ast$ and $\chi_\ast$.  We will now consider how the infationary
dynamics and predictions for $\zeta$ depend on these parameters.

\subsection*{Background trajectories}

\begin{figure}[h!]
\begin{subfigure}[t]{0.49\textwidth}
    \includegraphics[width = 0.99\columnwidth]{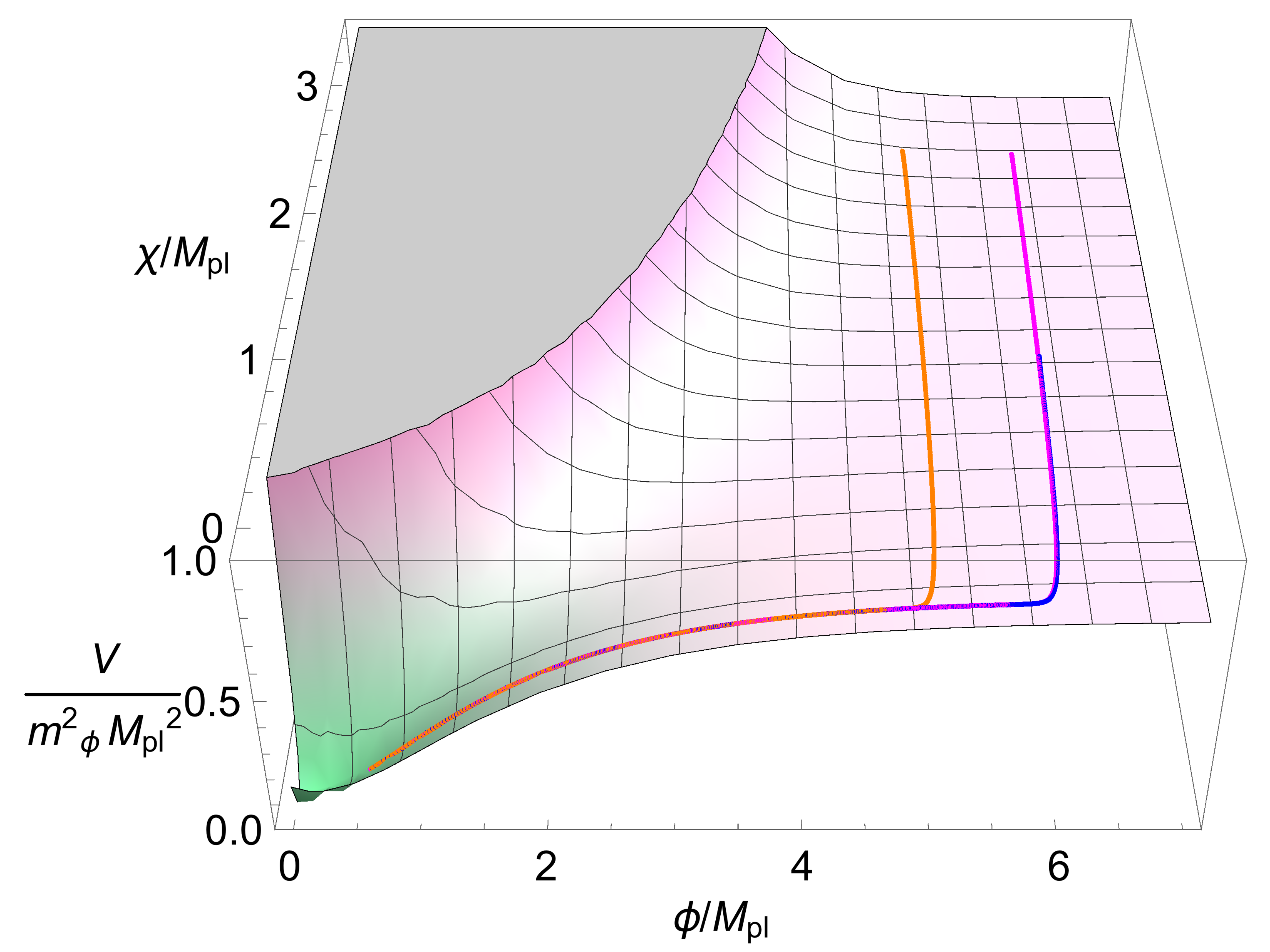} 
\caption{}
\end{subfigure}
\begin{subfigure}[t]{0.49\textwidth}
   \includegraphics[width = 0.99\columnwidth]{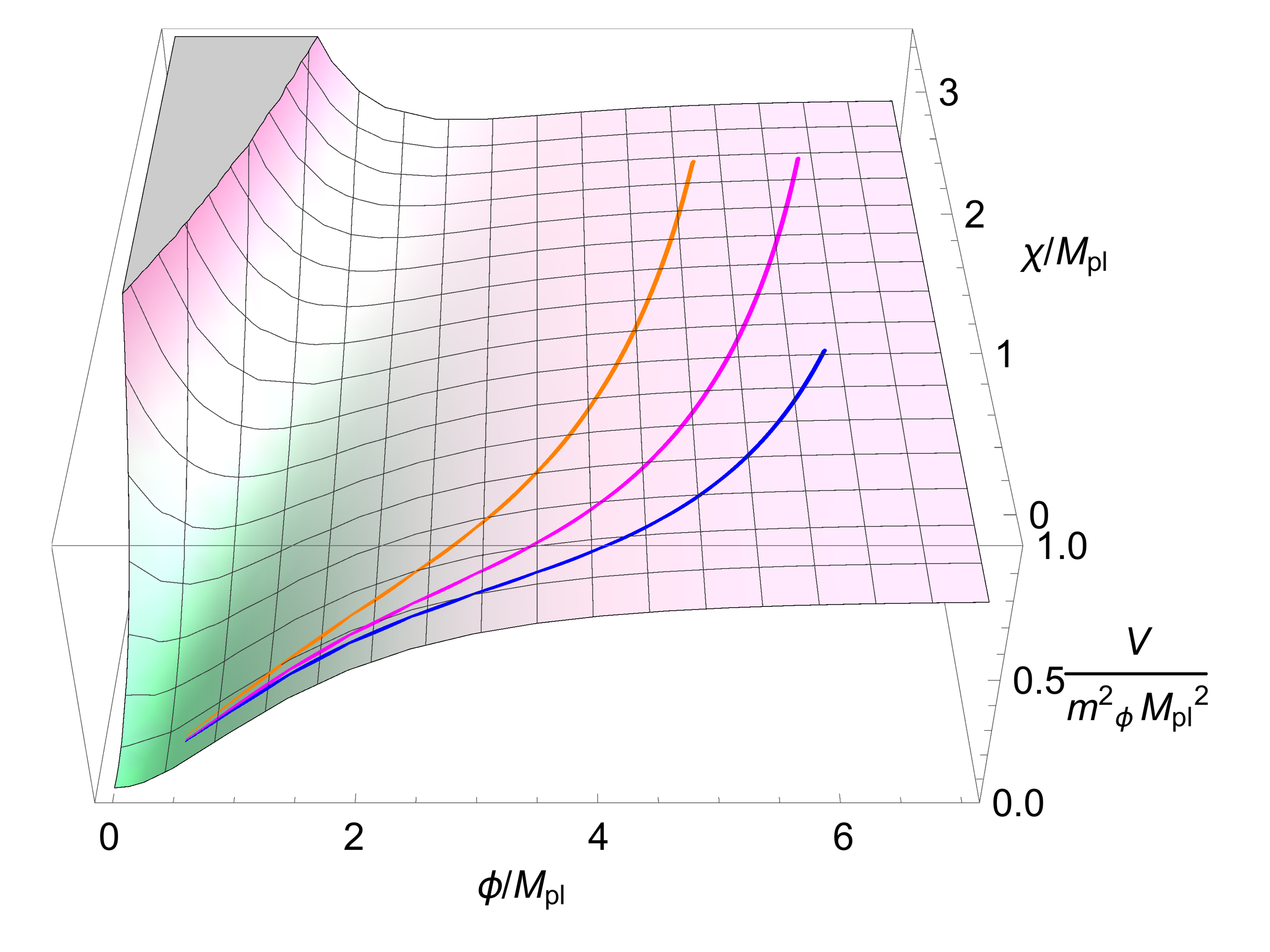} 
\caption{}
\end{subfigure}
\centering
\begin{subfigure}[t]{0.49\textwidth}
   \includegraphics[width = 0.99\columnwidth]{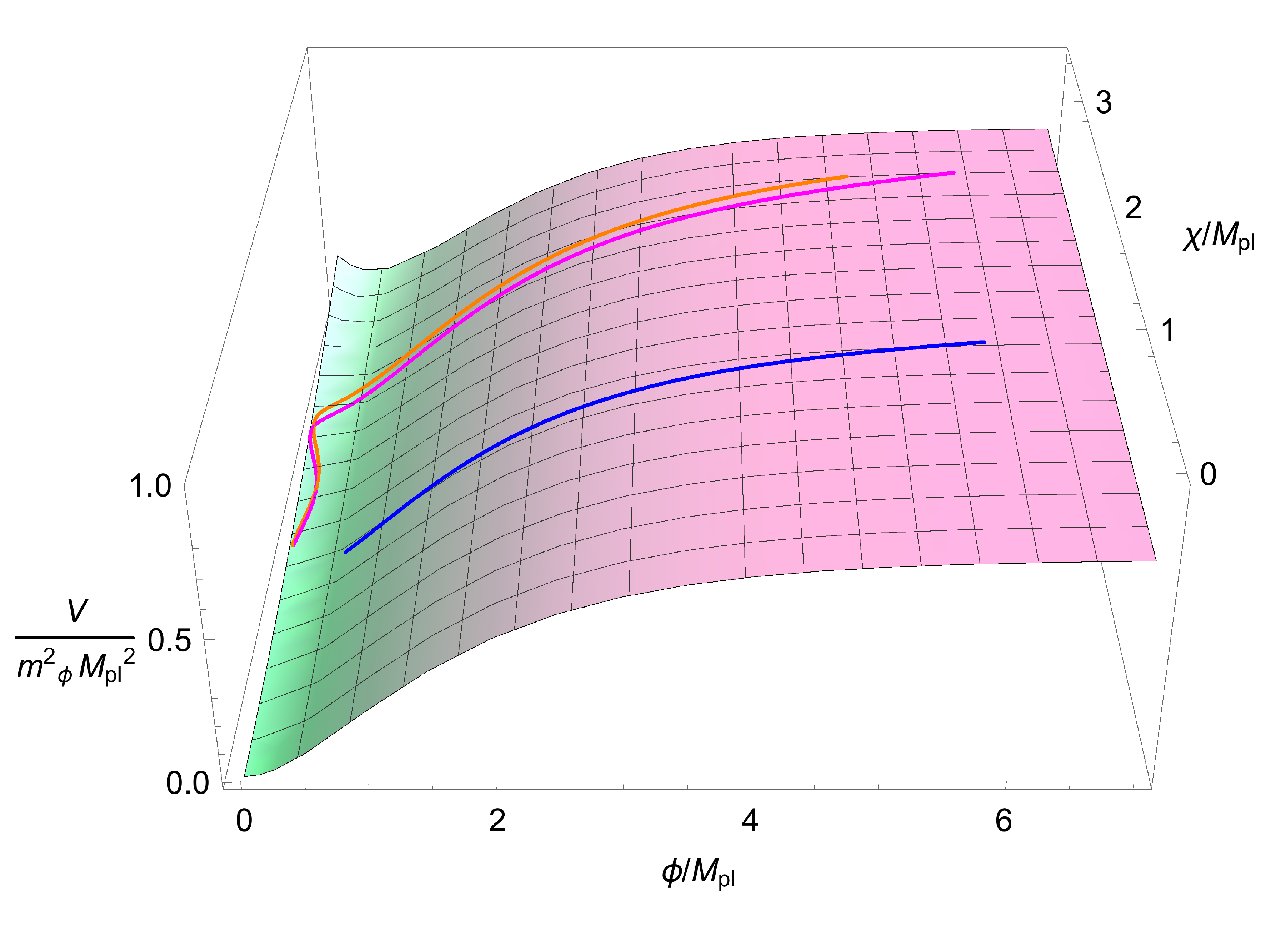} 
\caption{}
\end{subfigure}
 \caption{Examples trajectories for three different values of $R_{\rm
 mass}$. We show the cases a) $R_{\rm mass}=5.0$, b) $R_{\rm mass}=1.0$,
 and c) $R_{\rm mass}=0.2$. For each value of $R_{\rm mass}$ three
 trajectories are plotted, with the initial conditions given as
 $(\phi_*/M_{pl},\chi_*/M_{pl})=(6.0,3.0)$ (magenta line),
 $(\phi_*/M_{pl},\chi_*/M_{pl})=(5.0,3.0)$ (orange line) and
 $(\phi_*/M_{pl},\chi_*/M_{pl})=(6.0,1.5)$ (blue line).}  \label{trjs1}
\end{figure}

In light of the preceding discussion, we see that the shape of
trajectories in field space will be independent of $m_\phi$.  As such,
the only remaining parameters are $R_{\rm mass}$, $\phi_\ast$ and
$\chi_\ast$.  Broadly speaking, we are interested in the three regimes
$R_{\rm mass}>1$, $R_{\rm mass}\sim 1$ and $R_{\rm mass}<1$, and in
Fig.~\ref{trjs1} we plot example trajectories for the representative
values $R_{\rm mass} = 5,\,1,\,1/5$.  In each case we consider the three
sets of initial conditions $(\phi_\ast/M_{pl},\chi_\ast/M_{pl}) =
(6,3)$, $(\phi_\ast/M_{pl},\chi_\ast/M_{pl}) = (5,3)$ and
$(\phi_\ast/M_{pl},\chi_\ast/M_{pl}) = (6,1.5)$, and each trajectory is
evolved until inflation ends.  When interpreting the trajectories, one
has to be careful to recall that it is not only the potential shape that
is important, as the effect of the non-flat field space must also be
taken into account.  In this model, for example, we have $\mathcal
G^{\chi\chi} = e^{2\alpha\phi}$.  Given that the slow-roll equation of
motion for $\chi$ takes the form $3H\dot\chi \simeq -\mathcal
G^{\chi\chi}V_{,\chi}$, we can expect that for super-Planckian values of
$\phi$ the velocity is enhanced compared to what we would naively expect
from the gradient of the potential alone.  Nevertheless, the
trajectories in Fig.~\ref{trjs1} qualitatively agree with our naive
expectation.

In the case $R_{\rm mass} = 5$ we find that the trajectories first
rapidly evolve in the $\chi$ direction, with most of inflation then
taking place as the trajectory proceeds along the local minimum at $\chi
=0$.  Given that the potential reduces to the single-field $R^2$
potential at $\chi = 0$, we expect the last stage of inflation to be
indistinguishable from the original $R^2$ model.  At the level of
perturbations, as the trajectory evolves along the local minimum we
expect that $\zeta$ should be conserved and that isocurvature
perturbations will decay, such that an adiabatic limit is approached.
Recall that in the original $R^2$ inflation model approximately $60$
$e$-foldings of inflation are obtained by taking $\phi_\ast/M_{pl}\simeq 5.5$.
As such, in the large $R_{\rm mass}$ limit, if we take any set of
initial conditions with $\phi_\ast/M_{pl}\gtrsim 5.5$, the final stage of
inflation along $\chi = 0$ will constitute the whole observable part of
inflation, and we thus expect that predictions for $\zeta$ and its
statistical properties will be indistinguishable from the original $R^2$
model.  

In the case $R_{\rm mass} = 1$, the trajectories are less trivial, in
the sense that they continue to turn throughout the evolution.
Correspondingly, we expect that $\zeta$ will continue to evolve
throughout inflation.  It is in this parameter region that an adiabatic
limit may not be reached by the end of inflation, and $\zeta$ may
continue to evolve through the (p)reheating epoch.  If this is the case,
then the correlation functions of $\zeta$ that we find at the end of
inflation should not be directly compared with observations.

Finally, in the case $R_{\rm mass} = 1/5$, the trajectories are again as
expected, with essentially two stages of inflation taking place.
Initially the trajectories evolve in the $\phi$ direction, with the
potential profile in the $\phi$ direction being very similar to that of
the original $R^2$ inflation model.  In the cases of the orange and
magenta trajectories, they then turn and inflation proceeds as they
evolve essentially in the $\chi$ direction but while oscillating about
the local minimum located close to but not exactly at $\phi = 0$.  For
this choice of $R_{\rm mass}$, these trajectories do not appear to fully
relax to the bottom of the local minimum before the end of inflation,
and so we might expect that $\zeta$ is still evolving.  In the
case of the blue trajectory, due to the smaller initial position
$\chi_\ast/M_{pl} = 1.5$, we find that there is no second stage of inflation
driven by the $\chi$ field.

A feature that it is common to all choices of $R_{\rm mass}$ is that
for $\chi = 0$ both the potential and $V_{,\phi}$
reduce to those of $R^2$ inflation, while $V_{,\chi}
= 0$.  Consequently, trajectories with $\chi_\ast = 0$ will evolve
purely in the $\phi$ direction, and we expect that predictions for
$\zeta$ will coincide with those of $R^2$ inflation.

As we move to non-zero values of $\chi$, deviations from the $R^2$
potential depend on $R_{\rm mass}$, $\phi$ and $\chi$.  For
super-Planckian values of $\phi$ satisfying $2\alpha\phi\gg 1$, such
that $e^{-2\alpha\phi}\ll 1$, deviations from the $R^2$ potential are
suppressed by a factor of $e^{-4\alpha\phi}$, and will therefore be
negligible for sufficiently large values of $\phi$.  This feature can be
seen in all panels of Fig.~\ref{trjs1}.  In order for the $\chi$-field
contribution to the potential to dominate at some super-Planckian value
of $\phi$, one would require $R_{\rm
mass}^2\chi^2\gg3M_{pl}^2e^{4\alpha\phi}/2$, i.e.  $R_{\rm mass}$ or
$\chi$ must be very large.  However, in such a parameter region we find
that $\epsilon_\phi \equiv M_{pl}^2(V_{,\phi}/V)^2/2\simeq
8M_{pl}^2\alpha^2>1$, such that $\epsilon_V = \epsilon_\phi +
\epsilon_\chi>1$, where $\epsilon_\chi \equiv
M_{pl}^2e^{2\alpha\phi}(V_{,\chi}/V)^2/2 $, i.e. the slow-roll
approximation breaks down.

Note that although the potential reduces to $m_\chi^2\chi^2/2$ if we
take $\phi = 0$, due to the interaction term we do not have $V_{,\phi}
=0$ when $\phi = 0$.  As such, even if we start with $\phi_\ast = 0$, we do not
necessarily obtain chaotic inflation along the $\chi$ direction.
Indeed, $\dot\phi$ is positive along the axis $\phi = 0$, and if we
calculate $\epsilon_\phi$ we again find $\epsilon_\phi =
8M_{pl}^2\alpha^2>1$, such that $\epsilon_V>1$ and the slow-roll
approximation is violated.  Nevertheless, in the limit $R_{\rm mass }\ll
1$, with appropriate initial conditions we do find that the final stages
of inflation essentially coincide with quadratic chaotic inflation driven by the
$\chi$ field.  As can be seen in the third panel of Fig~\ref{trjs1}, for
small values of $R_{\rm mass}$ and sufficiently super-Planckian initial
conditions for $\phi$ and $\chi$, we obtain two stages of inflation.
The first stage is driven by $\phi$, and once $\phi$ reaches its minimum
the second stage is driven essentially by $\chi$.  The minimum of the potential in
the $\phi$ direction lies on the curve defined by
\begin{align}
\frac{2}{3}R_{\rm mass}^2\frac{\chi^2}{M_{pl}^2} = e^{2\alpha\phi}-1.\label{mindef}
\end{align}
As such, if $R_{\rm mass}$ is small enough to ensure that $R_{\rm
mass}\chi/M_{pl}\ll \sqrt{3/2}$, we find that the minimum lies very close
to the $\chi$ axis, with $\phi/M_{pl}\ll \sqrt{3/2}$.  In the same limit
$R_{\rm mass}\chi/M_{pl}\ll \sqrt{3/2}$, we find that along the minimum
with respect to $\phi$ the potential and its derivative with respect to
$\chi$ are approximately given as
\begin{align}
 \left.V\right|_{V_{,\phi}=0} \simeq \frac{1}{2}m_\chi^2\chi^2,
 \qquad \left. V_{,\chi}\right|_{V_{,\phi}=0} \simeq  m_\chi^2\chi,
\end{align}
i.e. they coincide with the case of a quadratic mass term for the $\chi$
field.  Consequently, in the limit $R_{\rm mass}\ll 1$, or more
precisely $R_{\rm mass}\ll \sqrt{3/2}M_{pl}/\chi_\ast$, once the $\phi$
field has evolved to its minimum we expect quadratic chaotic inflation
driven by $\chi$ to take place.  If we require that this stage of
chaotic inflation lasts for approximately $ 60$ $e$-foldings, then this
means we require $\chi_\ast \simeq 15.5 M_{pl}$, which in turn gives the
condition $R_{\rm mass} \ll \sqrt{3/2}/15.5 \simeq 0.08$.  In analogy
with the large $R_{\rm mass}$ limit, we find that for $R_{\rm mass }\ll
0.08$ the whole of the observable period of inflation will essentially
coincide with quadratic chaotic inflation driven by $\chi$ if we take
any initial conditions with $\chi_\ast\gtrsim 15.5 M_{pl}$.  At the level of
perturbations, in analogy with the large $R_{\rm mass}$ case, as the
trajectory evolves along the the local minimum close to $\phi = 0$ we
expect that $\zeta$ will be conserved and that isocurvature
perturbations decay, such that an adiabatic limit is approached.

    \begin{figure}[h!]
     \begin{center}
      \includegraphics[width=10cm,bb=0 0 1024 768]{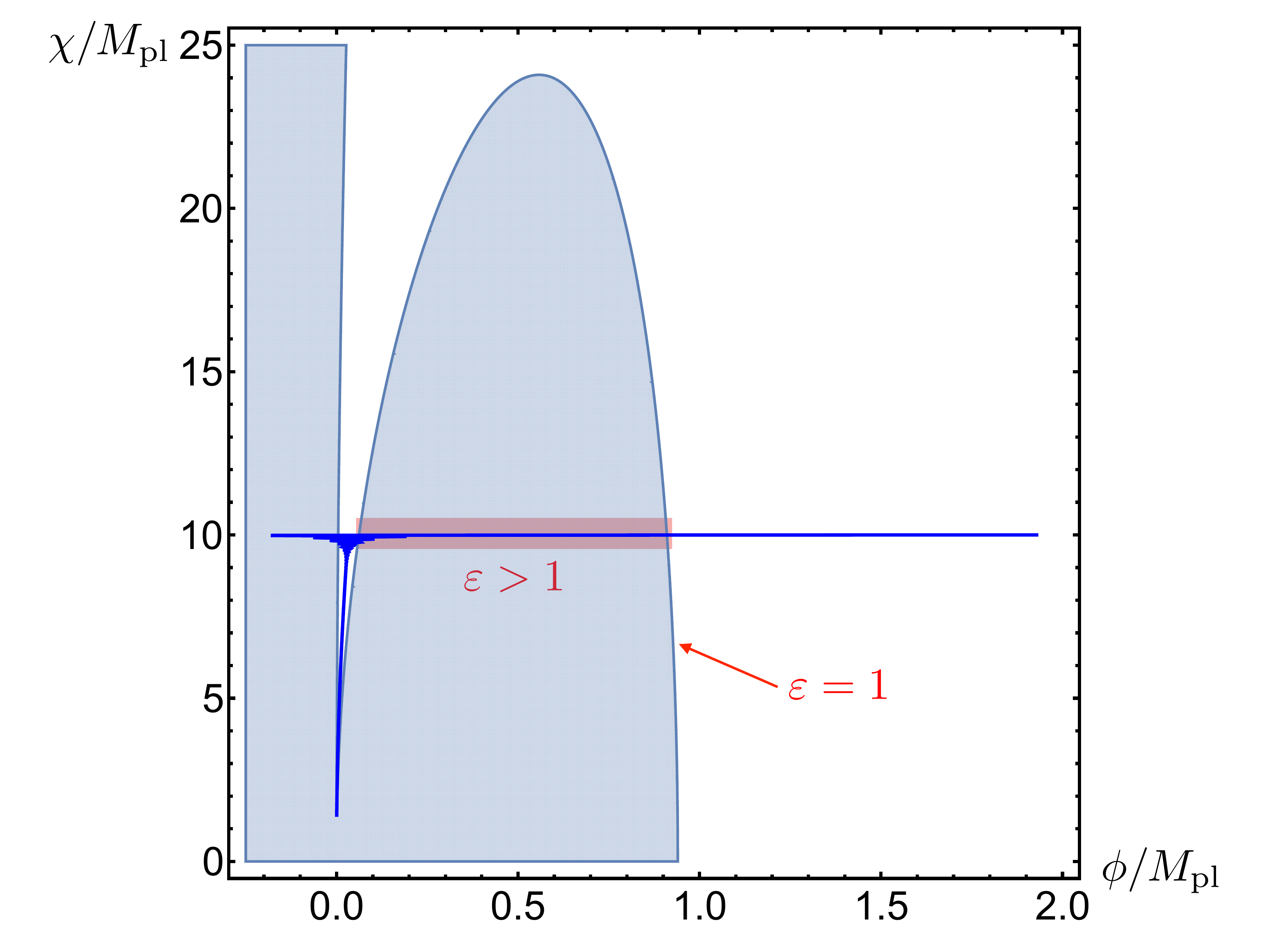} 
      \caption{An example trajectory for
      $R_{\rm mass}=0.02$ and
      $(\phi_\ast/M_{pl},\chi_\ast/M_{pl})=(2.0,10.0)$. The
      region with $\epsilon_V>1$ is shaded in blue.  The trajectory
      consists of two inflationary stages separated by a
      non-inflationary stage.}
      \label{fast_roll}
     \end{center}
\end{figure}
    

Given that the mass ratio $R_{\rm mass}= 0.2$ considered in
Fig.~\ref{trjs1} is not so small, in Fig.~\ref{fast_roll} we plot an
example trajectory for the case $R_{\rm mass} = 0.02$ and the initial
conditions $(\phi_\ast/M_{pl},\chi_\ast/M_{pl})=(2.0,10.0)$ . 
We also show the region where $\epsilon_V>1$.
Similar to the case $R_{\rm mass} = 0.2$ considered in Fig.~\ref{trjs1},
the trajectory first evolves in the $\phi$ direction and the potential
profile essentially coincides with $R^2$ inflation.  Intermediately,
when $\phi$ drops below $M_{pl}$, one thus finds that $\epsilon_V$
becomes greater than unity and inflation temporarily ceases.  However,
once the trajectory reaches the local minimum, $\epsilon_V$ once again
becomes less than unity and inflation recommences, with the subsequent
trajectory evolving essentially in the $\chi$ direction.  It is thus
important when considering small values of $R_{\rm mass}$ that we do not
terminate our integration of the trajectory prematurely, in order not to
miss the second stage of inflation.  Note that there is a period during
which the $\phi$ field oscillates about its minimum, and during this
period one might expect the $\phi$ field to decay into any matter fields
to which it is coupled, including the $\chi$ field.  Indeed, due to the
non-minimal coupling of $\phi$ to the Ricci scalar in the Jordan frame,
we expect there to at least be gravitationally induced couplings between
$\phi$ and any other matter fields present, see
e.g.~\cite{Faulkner:2006ub,Watanabe:2006ku}.  However, in the following
we neglect the possible decay of the $\phi$ field, postponing a careful
consideration of this effect to future work.

\subsection*{Evolution of perturbations}

Having given some example background trajectories, we now consider the
evolution of $\zeta$, or more precisely its correlation functions.  As
discussed above, in single field inflation we know that $\zeta$ is
conserved on superhorizon scales, while in multi-field inflation it is
sourced by isocurvature perturbations if the trajectory in field space
deviates from a geodesic, see e.g.~\cite{Gordon:2000hv,Langlois:2008mn}.

    \begin{figure}[h!]
     \begin{center}
      \includegraphics[width=10cm,bb=0 0 1024 768]{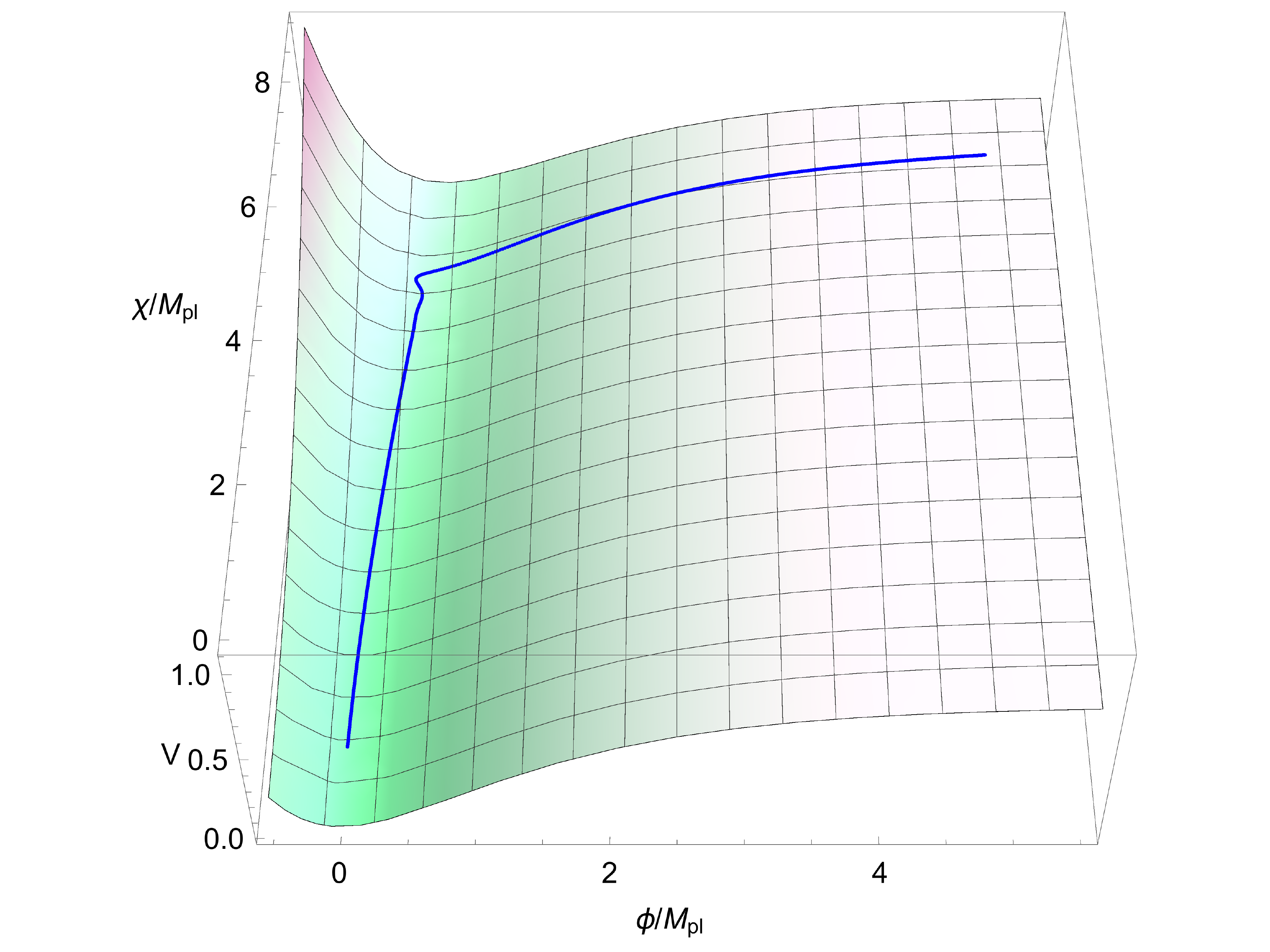} 
      \caption{An example trajectory with $R_{\rm mass}=0.1$ and initial
      conditions $(\phi_*/M_{pl},\chi_*/M_{pl})=(5.0,8.0)$.  Inflation lasts for a
      total of $58$ $e$-foldings.}
      \label{evol_trj}
     \end{center}
\end{figure}

\begin{figure}[h!]
     \begin{subfigure}[t]{0.49\textwidth}
      \includegraphics[width=0.99\columnwidth]{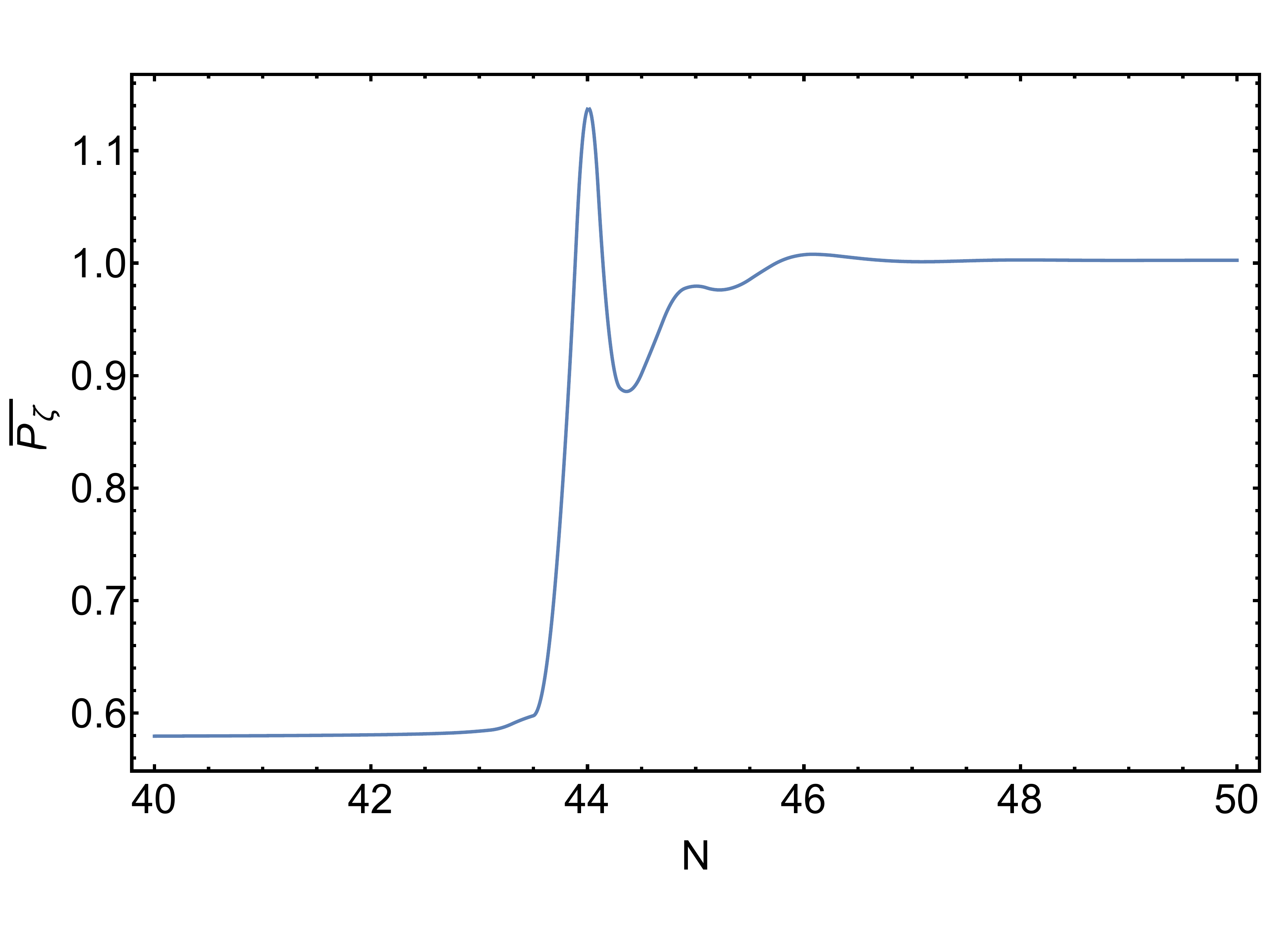}
     \end{subfigure}
      \begin{subfigure}[t]{0.49\textwidth}
      \includegraphics[width=0.99\columnwidth]{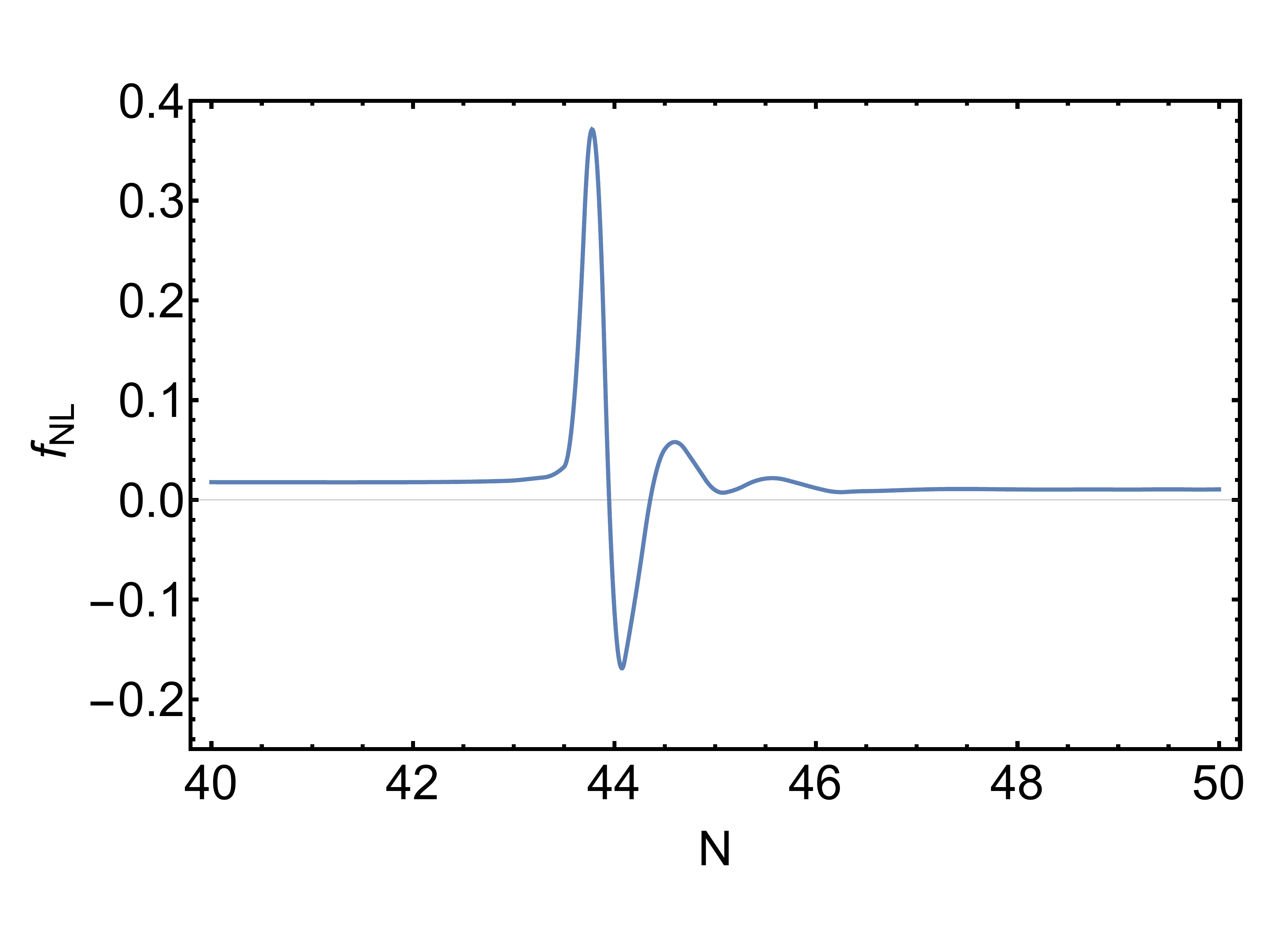} 
      \end{subfigure}
 \caption{Evolution of the normalised power spectrum $\overline
        P_{\zeta} = P_{\zeta}(N)/ P_{\zeta}(N_{\rm final})$ (left panel)
        and $f_{NL}$ (right panel) for the example trajectory plotted in
        Fig.~\ref{evol_trj}.} \label{evol_PsfNL}
\end{figure}

Perhaps the most interesting evolution of $\zeta$ and its correlation
functions is observed in the case of small $R_{\rm mass}$.  As an
example, we consider the parameters $R_{\rm mass} = 0.1$ and
$(\phi_\ast/M_{pl},\chi_\ast/M_{pl}) = (5,8)$.  The background trajectory for this
choice of parameters is shown in Fig.~\ref{evol_trj}.  Given that
$R_{\rm mass }\ll 1$, we see that the trajectory first evolves in the
$\phi$ direction, before moving along the local minimum that runs almost
parallel to the $\chi$ axis.  In total there are approximately $60$
$e$-foldings of inflation, with the turn occurring at $N\sim 45$.  In
the left panel of Fig.~\ref{evol_PsfNL} we plot the evolution of the
power spectrum, normalised by the final value.  As expected, it remains
constant for the first $40$ $e$-folds, and is then sourced by
isocurvature modes as the trajectory turns at around $N\sim 45$.  Given
the relatively large mass hierarchy, $\mathcal P_\zeta$ is seen to oscillate
as the trajectory oscillates about the local minimum, before again
approaching a constant as an essentially single-field limit is reached.

In the right panel of Fig.~\ref{evol_PsfNL} we show the evolution of
$f_{\rm NL}$ for the same trajectory.  Up until the turn it is
negligibly small, with $f_{NL}\sim\mathcal O(10^{-2})$.  During the turn
and subsequent oscillations we find that $f_{NL}$ also oscillates, with
a peak amplitude of $f_{NL}\simeq 0.35$.  In the final stage, however,
$f_{NL}$ relaxes back down to an unobservably small value of $ \mathcal
O(10^{-2})$.  The behaviour of the power spectrum and $f_{NL}$ in
this example are qualitatively very similar to that observed in double
quadratic inflation models, see
e.g. \cite{Vernizzi:2006ve,Yokoyama:2007dw,Watanabe:2011sm} and references therein.

We have also considered the evolution of $\mathcal P_\zeta$ and $f_{NL}$
in the other regimes $R_{\rm mass} > 1$ and $R_{\rm mass}\sim 1$.  For
trajectories with $R_{\rm mass} >1$, such as those shown in the first
panel of Fig.~\ref{trjs1}, due to the fact that the trajectories quickly
evolve to the $\phi$ axis and reach an effectively single-field
trajectory along the $\phi$ axis, we find that $\mathcal P_\zeta$ and
$f_{NL}$ also quickly reach constant values, with $f_{NL}\sim \mathcal
O(10^{-2})$.  Note that the background trajectories shown in the first
panel of Fig.~\ref{trjs1} do not oscillate about the $\phi$ axis, and
correspondingly we find that $\mathcal P_\zeta$ and $f_{NL}$ also do not
oscillate before settling to their constant values.  For trajectories
with $R_{\rm mass }\sim 1$, such as those shown in the second panel of
Fig.~\ref{trjs1}, we find that the evolution of $\mathcal P_\zeta$ and
$f_{NL}$ is much more gradual, with $f_{NL}$ remaining $\mathcal
O(10^{-2})$ throughout the evolution.

\subsection*{Exploring and constraining parameter space}

\begin{figure}[h!]
 \begin{center}
  \includegraphics[width=0.95\textwidth]{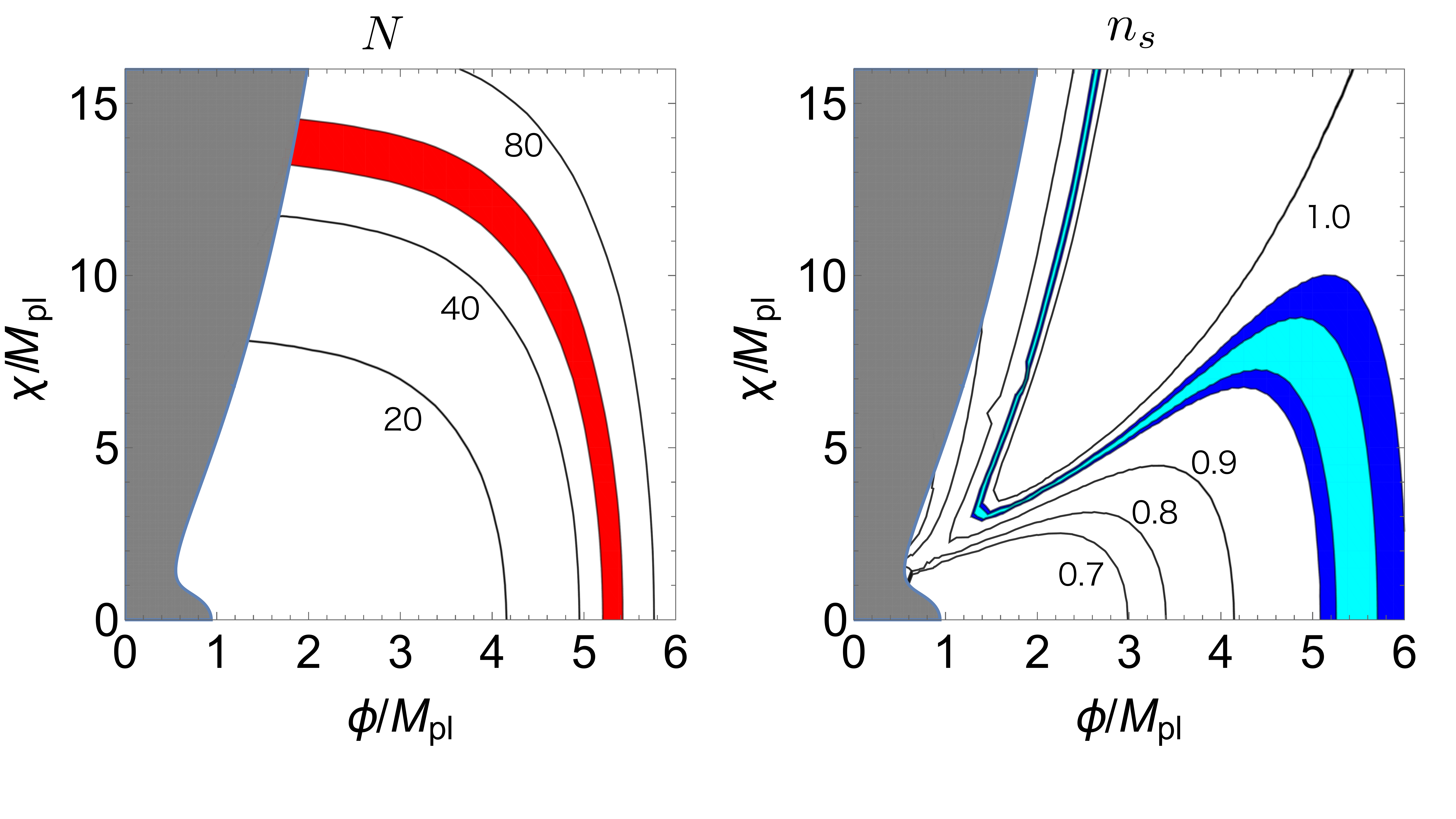}
  \includegraphics[width=0.95\textwidth]{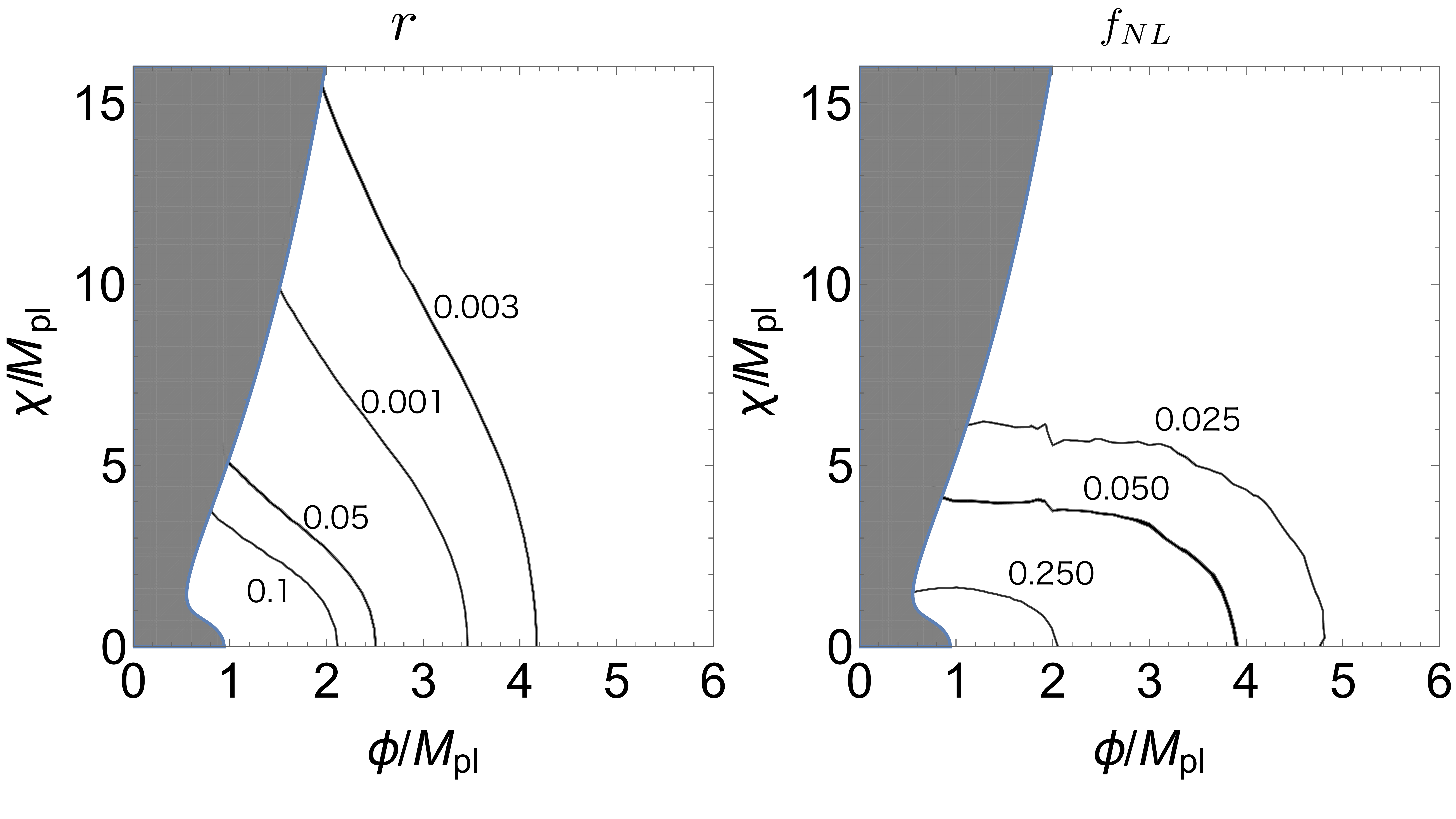}
  \caption{Predictions in the $\phi_\ast$--$\chi_\ast$ plane for the
  $e$-folding number $N$ (upper left), spectral tilt $n_s$ (upper
  right), tensor-to-scalar ratio $r$ (lower left) and non-Gaussianity
  parameter $f_{\rm NL}$ (lower right) for the case $R_{\rm mass}=1.0$.
  The red shaded region in the upper left plot shows the initial
  conditions for which $50<N<60$. The light blue (dark blue) shaded
  region in the upper right plot indicates the range of initial
  conditions for which $n_s$ lies within 1-$\sigma$ (2-$\sigma$) of the
  observed value.  The grey shaded region in all plots corresponds to
  where $\epsilon_V> 1$.}  \label{N_ns_r_fNL}
 \end{center}
\end{figure}

Having looked at representative example trajectories in the three
regimes $R_{\rm mass}<1$, $R_{\rm mass }\sim 1$ and $R_{\rm mass}>1$, we
now proceed to put constraints on the parameters $m_\phi$ and $R_{\rm
mass}$.  In doing so we consider thirty-one different mass ratios in the
range $10^{-3}\leq R_{\rm mass} \leq 10^{3}$, distributed evenly over
$\log R_{\rm mass}$.  For each value of $R_{\rm mass}$ we then consider
a $50\times 50$ grid of initial conditions $(\phi_\ast,\chi_\ast)$, with
$\phi_\ast$ spanning the range $0\le\phi_\ast\le 6M_{pl}$ and $\chi_\ast$
spanning the range $0\le\chi_\ast\le 16 M_{pl}$.\footnote{From our knowledge of
the $R^2$ and quadratic chaotic inflation models, we
know that taking $\phi_\ast>6$ or $\chi_\ast>16$ will always give
$N>60$, but observationally we are only interested in the last $60$
$e$-folds of inflation.  Hence our choice of maximum $\phi_\ast$ and
$\chi_\ast$.}  Next we neglect any points on the grid for which either
$\epsilon_V>1$ or $|\eta_V|>1$, i.e. we require that the slow-roll
approximation is valid at the horizon-crossing time.  For the remaining
points we are then able to determine $N$, $\mathcal P_\zeta/m_\phi^2$,
$n_s$, $r$ and $f_{NL}$ without needing to specify $m_\phi$.  Neglecting
points that do not give $50<N<60$, for each of the remaining points we
perform a chi-squared analysis to determine the range of $m_\phi$ for
which the predictions for $\mathcal P_\zeta$, $n_s$, $r$ and $f_{NL}$
lie within 1- and 2-$\sigma$ of the observed values summarised at the
end of Sec.~\ref{PSBS}.  At this point, for every observationally
allowed set of initial conditions $(\phi_\ast,\chi_\ast)$ we have a
maximum and minimum allowed $m_\phi$.  To find the overall maximum and
minimum allowed values of $m_\phi$ for a given $R_{\rm mass}$, we must
then take the maximum of all the maxima and the minimum of all the
minima.  Note that for any value of $R_{\rm mass}$ we are guaranteed to
find a non-vanishing allowed range of $m_\phi$, as we will always
recover the predictions of $R^2$ inflation if we take
$\chi_\ast=0$.

\begin{figure}[t]
 \begin{center}
  \includegraphics[width=10.5cm,bb=0 0 1024 768]{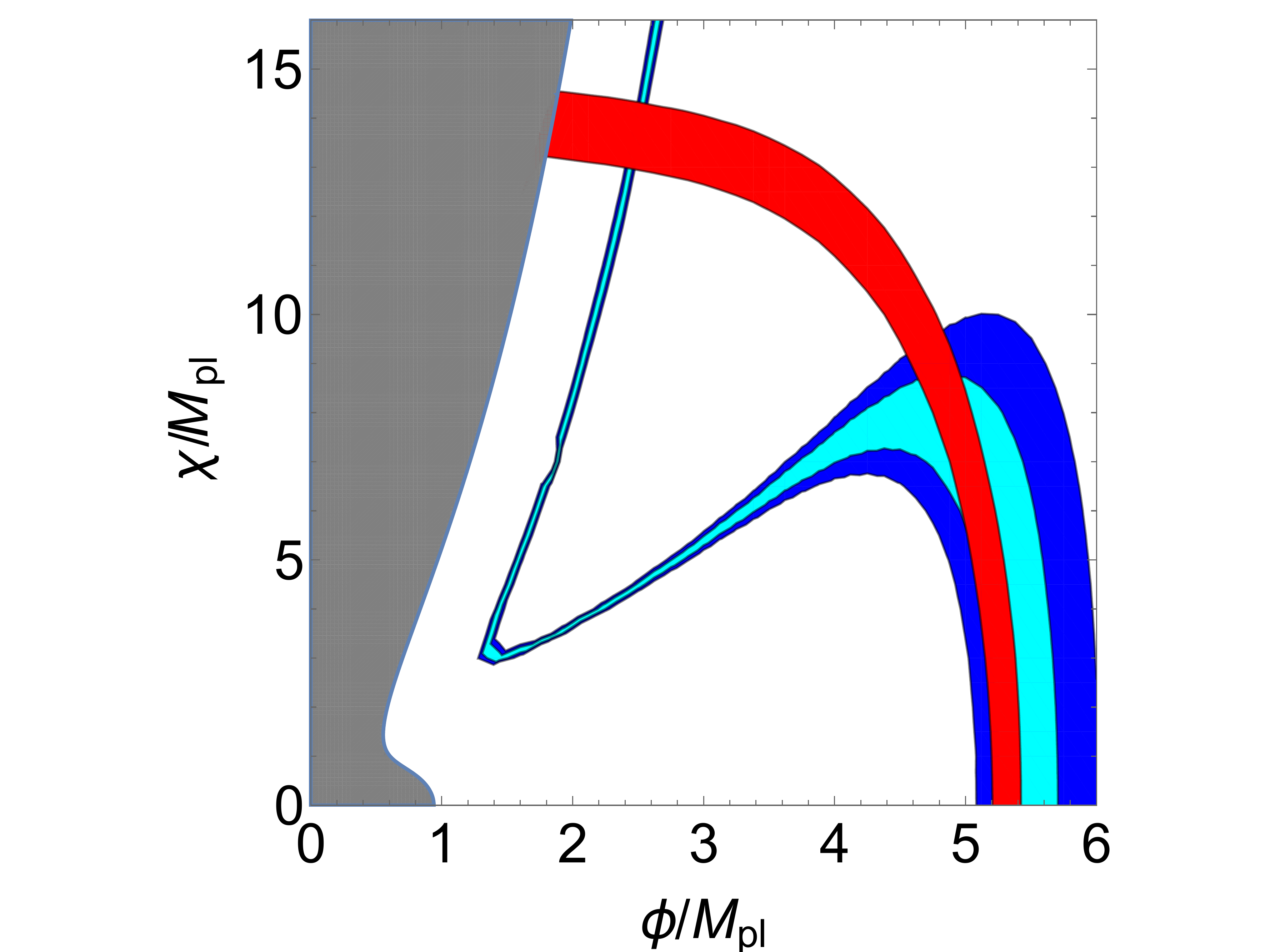}
  \caption{Combined constraints in the $\phi_\ast$--$\chi_\ast$ plane
  for the case $R_{\rm mass} = 1$.  The red shaded region corresponds to
  the constraint $50<N<60$, while the light- and dark-blue regions
  correspond to the 1- and 2-$\sigma$ observational constraints on $n_s$
  given at the end of Sec.~\ref{PSBS}.  Predictions for $r$ and $f_{\rm
  NL}$ are consistent with observations for all sets of initial
  conditions.  The grey shaded region corresponds to where $\epsilon_V>1$.}
  \label{reg_N_ns}
 \end{center}
\end{figure}

As an example, in Fig.~\ref{N_ns_r_fNL} we show the predictions for $N$,
$n_s$, $r$ and $f_{NL}$ in the $\phi_\ast$--$\chi_\ast$ plane for the
case $R_{\rm mass } = 1$.  In the plot of $N$ we highlight in red the
region for which $50<N<60$.  Similarly, in the plot of $n_s$ we
highlight in light- and dark-blue the regions that fall within $1$- and
$2$-$\sigma$ of the observed value.  For all values of $\phi_\ast$ and
$\chi_\ast $ we find that $r$ and $f_{NL}$ are consistent with
observational constraints.  In Fig.~\ref{reg_N_ns} we combine the
constraints coming from $N$ and $n_s$, which allows us to determine the
region in the $\phi_\ast$--$\chi_\ast$ plane in which the horizon exit
point must lie.  The intersection of the red shaded region with the
$\phi$ axis corresponds to the initial conditions for $R^2$ inflation.
As we move away from the $\phi$ axis we see that there is quite an
extended region that remains in agreement with observations.
Interestingly, there is another small allowed region towards the top
left-hand corner of the $\phi_\ast$--$\chi_\ast$ plane.

\begin{figure}[h!]
 \begin{center}
  \includegraphics[width=11cm,bb=0 0 1024 768]{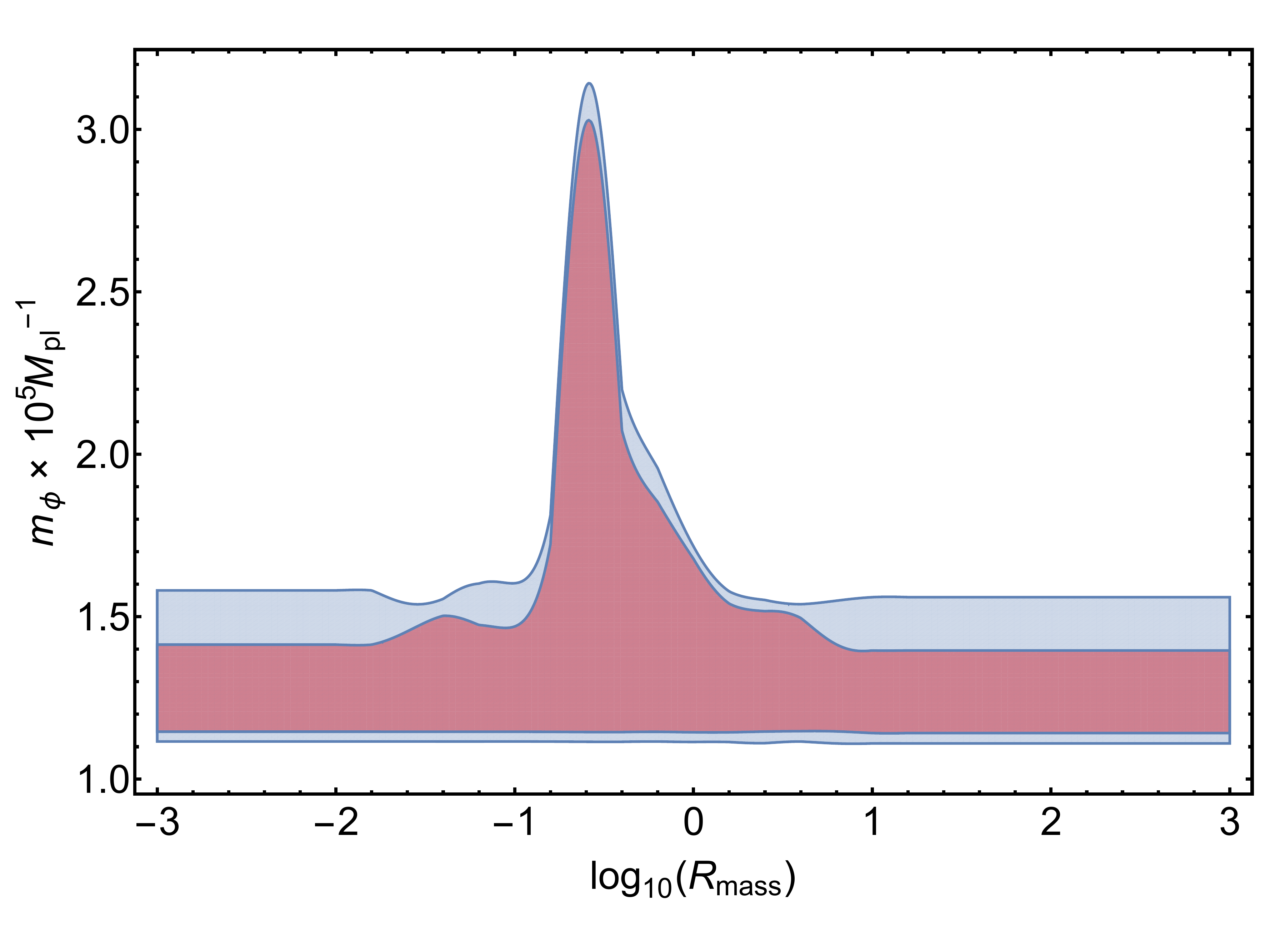}
  \caption{Allowed regions in the $R_{\rm mass}$--$m_{\phi}$
  plane at $1$-$\sigma$ (red shaded region) and
  $2$-$\sigma$ (blue shaded region).} \label{mphi}
 \end{center}
\end{figure}

The obtained constraints on $m_{\phi}$ as a function of $R_{\rm mass}$
are shown in Fig.~\ref{mphi}. In the limits of both small and large
$R_{\rm mass}$ we find that the allowed range is consistent with that of
$R^2$ inflation, for which slow-roll estimates give
$m_{\phi}\simeq(1.2$--$1.4) \times 10^{-5}M_{pl}$ for $N = 50$--$60$.
In the large $R_{\rm mass}$ limit this has a
relatively simple interpretation. As the $\chi$ field becomes more
massive one approaches a limit in which all slow-roll trajectories
satisfiying $\epsilon_V,|\eta_V|\ll 1$ at horizon crossing and giving
$50<N<60$ correspond to effecitvely single-field trajectories that
evolve along the $\phi$ axis, where the potential reduces to that of
$R^2$ inflation.  We can see from Fig.~\ref{mphi} that such a limit is
reached for $R_{\rm mass}\gtrsim 10$.  In the small $R_{\rm mass}$ limit
the situation is less clear.  The fact that the allowed range of
$m_\phi$ approaches a constant can be understood as follows. So long as
$R_{\rm mass}$ is smaller than some critical value --- which our results
suggest is around $R_{\rm mass}\simeq 10^{-2}$ --- one finds that for a
given set of initial conditions the last $60$ $e$-foldings of inflation is
well approximated by a stage of $R^2$ inflation followed by a stage of
quadratic chaotic inflation, as was observed in Fig.~\ref{fast_roll}.
The fact that the allowed range of $m_\phi$ coincides with that of $R^2$
inflation, however, is not so obvious.  As $\chi_\ast$ is increased from
$0$ to $16 M_{pl}$ (and $\phi_\ast$ is correspondingly adjusted to give the
desired number of $e$-foldings), we expect that predictions for
$\mathcal P_\zeta$, $r$, $n_s$ and $f_{NL}$ will interpolate between
those of $R^2$ inflation and those of quadratic chaotic inflation.
While the latter are ruled out by observations, one might naively expect
that intermediately there are values of $\chi_\ast$ that give
predictions deviating from $R^2$ inflation but still in agreement with
observations, which in turn would naively alter the allowed range of
$m_\phi$.  However, our results suggest that the allowed range of
$m_\phi$ is essentially unaffected.

For intermediate values of $R_{\rm mass}$ the obtained bounds on
$m_\phi$ are found to deviate from those of $R^2$ inflation. As we can
see in Fig.~\ref{mphi}, the allowed range of $m_{\phi}$ has a peak of
$m_\phi\simeq 3\times 10^{-5}M_{pl}$ at around $\log_{10}\left(R_{\rm
mass}\right)\simeq -0.5$ ($R_{\rm mass} \simeq 0.3$).  However, one must
bear in mind that for some of the trajectories in this parameter range
an adiabatic limit will not have been reached by the end of inflation.
As such, it may be that the constraints in this region would change if
effects of the (p)reheating epoch were taken into account.

\begin{figure}[h!]
 \begin{subfigure}[t]{0.49\textwidth}
  \includegraphics[width=0.99\columnwidth]{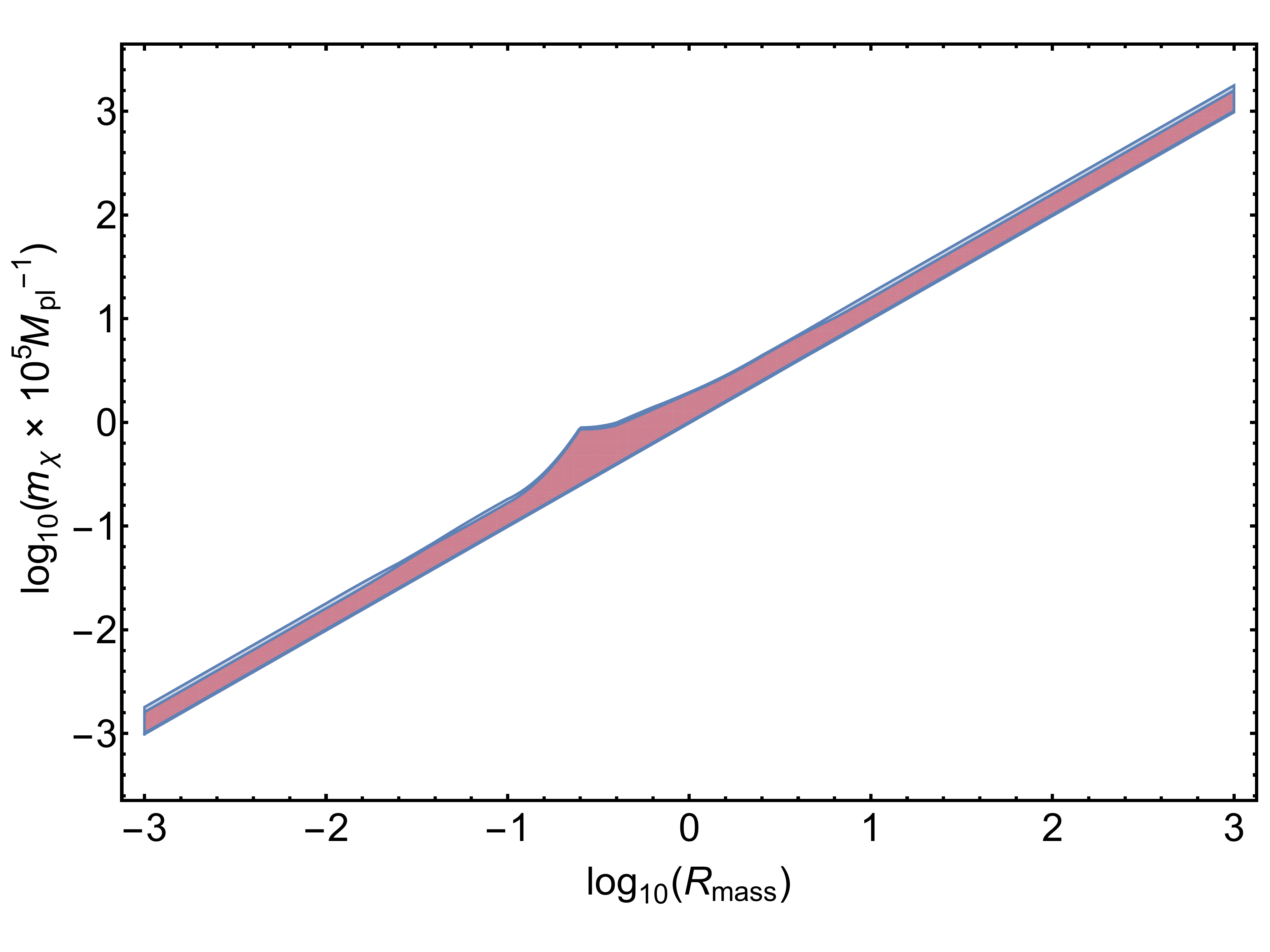} 
  \caption{}
  \label{mchi}
 \end{subfigure}
 \begin{subfigure}[t]{0.49\textwidth}
  \includegraphics[width=0.99\columnwidth]{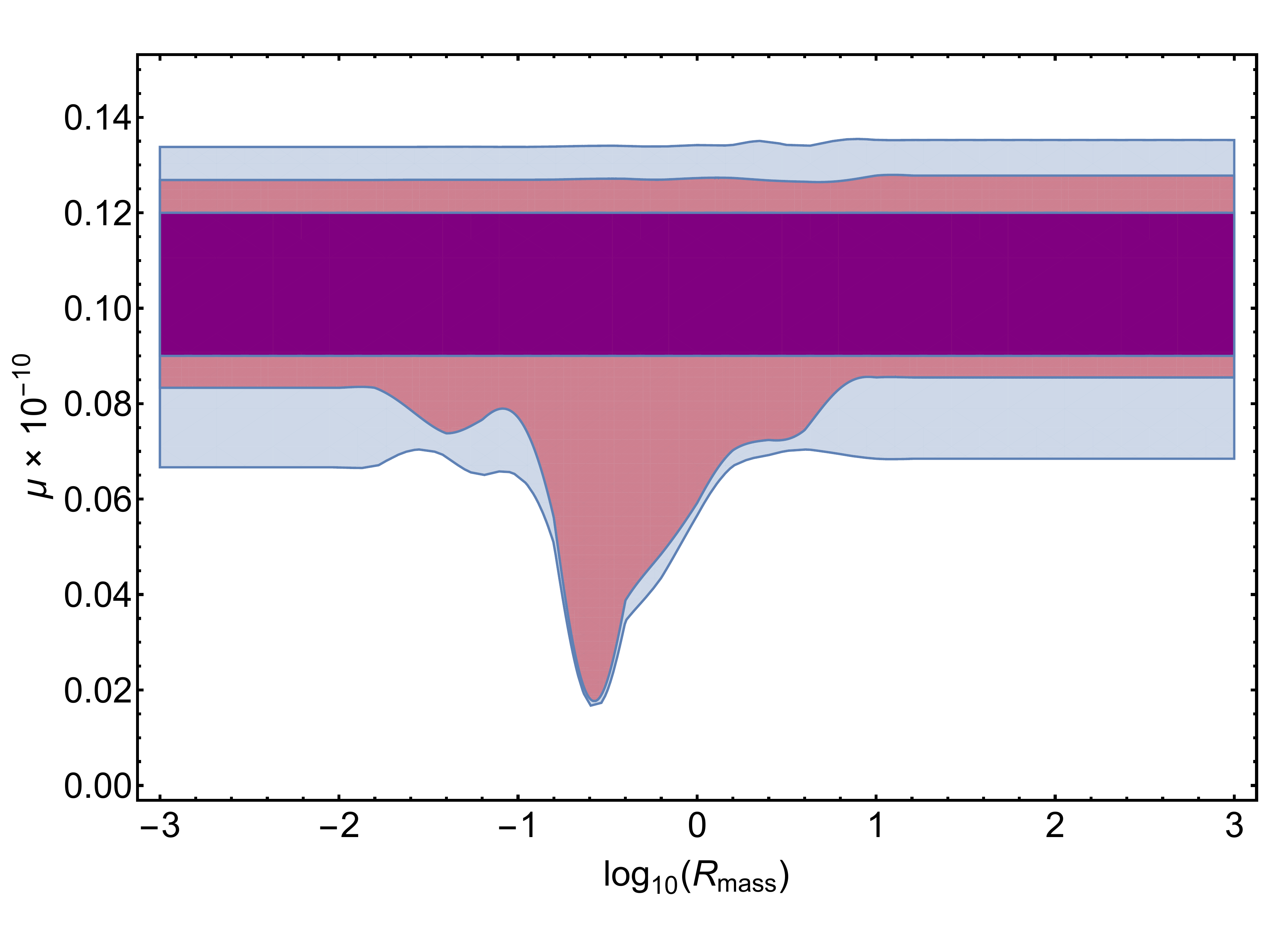}
  \caption{}  \label{mu}
 \end{subfigure}
\caption{(a) Allowed regions in the $R_{\rm mass}$--$m_{\chi}$ plane at
  $1$-$\sigma$ (red shaded region) and $2$-$\sigma$ (blue shaded
  region).  (b) Allowed regions in the $R_{\rm mass}$--$\mu$ plane at $1$-$\sigma$
  (red shaded region) and $2$-$\sigma$ (blue shaded region). The purple-shaded
  region shows $\mu=(0.9$--$1.2)\times10^{9}$, which corresponding to the case of
  the original $R^2$ inflation model with $N=$~50--60.}
\end{figure}

Using the definition of $R_{\rm mass}$, we can use the above constraints
on $m_\phi$ to also put bounds on $m_{\chi}$ as a function of $R_{\rm
mass}$.  The results are shown in Fig.~\ref{mchi}.  Similarly, recall
that in the Jordan frame representation of this model one has the
parameter $\mu$ instead of $m_\phi$, see eq.~\eqref{eqR_x1}.  Given that
these two parameters are related as $m^2_{\phi}=M^2_{pl}/(6\mu)$, we can
re-express our constraints on $m_\phi$ as constraints on $\mu$, and the
results are shown in Fig.~\ref{mu}. In the original $R^2$ inflation
model, slow-roll estimates determine that in order to satisfy
observational constraints one requires $\mu\simeq (0.9$--$1.2)\times 10^{9}$ for
$N$=50--60~\cite{Calmet:2016fsr, Netto:2015cba}, which is consistent with our
constraints.  In the multi-field extension of $R^2$ inflation
that we have considered, we find the allowed range of $\mu$ to be $\mu
\simeq (0.2$--$1.3)\times 10^{9}$. 

%
%
\section{Summary}
\label{summary}

In this paper we have considered a two-field inflation model based on a
simple multi-field extension of $R^2$ inflation.  In addition to a term
proportional to $R^2$, the Jordan frame action contains a canonical
scalar field $\chi$ with quadratic mass term. On re-writing the model as
a scalar-tensor theory and making a conformal transformation into the
Einstein frame, the model takes the form of a two-field inflation model
with a non-flat field space as shown in eq.~\eqref{Eaction}.  The first
field, $\phi$, corresponds the additional degree of freedom associated
with the $R^2$ term in the original action and is often referred to as
the scalaron.  This field has a canonical kinetic term and its potential
takes on the same form as in the original $R^2$ inflation model,
approaching a constant for super-Planckian values of $\phi$.  The second
field, $\chi$, on the other hand, has a non-canonical kinetic term that
depends exponentially on $\phi$, and its quadratic mass term similarly
contains an exponential coupling with $\phi$.

Assuming that the slow-roll approximation is valid at horizon crossing,
such that only the initial field positions have to be specified in
solving for the inflationary dynamics, the four free parameters of the
model are $m_\phi$, $R_{\rm mass} = m_\chi/m_\phi$, $\phi_\ast$ and
$\chi_\ast$.  In Sec.~\ref{Num} we have explored how the inflationary
dynamics and predictions for the correlation functions of $\zeta$ depend
on these four parameters, both qualitatively and quantitatively.

For $R_{\rm mass}\gtrsim 10$ we find that all slow-roll trajectories
satisfying $\epsilon_V,|\eta_V|\ll 1$ at horizon crossing and giving
$50<N<60$ follow an effectively single-field trajectory evolving along
the local minimum of the potential at $\chi = 0$.  Given that the
potential coincides with that of the original $R^2$ inflation model
along $\chi = 0$, the predictions for $\zeta$ and its correlation
functions also coincide with the original model.  In this region of
parameter space observational constraints give $m_\phi \simeq
(1.1$--$1.6)\times 10^{-5}M_{pl}$ at $2$-$\sigma$.
   
For $R_{\rm mass}\lesssim 10^{-2}$ we find that inflationary
trajectories consist of a stage of $R^2$ inflation driven by $\phi$
followed by a stage of quadratic chaotic inflation driven by $\chi$.
How the last observable $60$ $e$-foldings of inflation are divided
between these two stages depends on the initial conditions, and the
predictions for $\zeta$ and its correlation functions thus range from
those of $R^2$ inflation to those of quadratic chaotic inflation.
Interestingly, however, we find that the final constraints on $m_\phi$
are very similar to those obtained in the large $R_{\rm mass} $ limit,
namely they essentially coincide with the limits on $m_\phi$ obtained in
the original $R^2$ model.

Finally, in the parameter region $R_{\rm mass }\sim 1$, we find that the
constraints on $m_\phi$ are less tight, with $m_\phi \simeq
(1.1$--$3.2)\times 10^{-5}M_{pl}$ at $2$-$\sigma$.  Here it is less easy
to interpret the results, as the inflationary trajectories are truly
multi-field in nature, with both $\phi$ and $\chi$ evolving throughout
inflation in many cases and the dynamics very much depending on the
initial conditions.  Nevertheless, one can see that the net result of
these multi-field effects is to increase the allowed range of $m_\phi$
as compared to the original $R^2$ model, and in particular to
allow for larger values of $m_\phi$.

One issue that we have not fully addressed in this paper is the
possibility that $\zeta$ may continue to evolve after the end of
inflation. If an adiabatic limit is not reached by the end of inflation
then one should continue to follow the evolution of $\zeta$ through
(p)reheating and until an adiabatic limit is reached.  It is only the
final $\zeta$ that should then be compared with observations.  In the
cases $R_{\rm mass}\gg1$ and $R_{\rm mass}\ll 1$ the issue is naively
not important, as we expect an adiabatic limit to be reached before the
end of inflation for most inflationary trajectories.  For $R_{\rm
mass}\sim 1$, however, this is no longer the case, and so
post-inflationary evolution of $\zeta$ may affect the constraints on
$m_\phi$ in this region.  We hope to address this issue in future work.

%
\acknowledgments
We would like to thank M. Kar\v{c}iauskas for valuable discussions in the
early stage of this work. This work was partially supported by JSPS
Core-to-Core Program, Advanced Research Networks (T.M.), JSPS KAKENHI Grant Nos. 26247042, and JP1701131 (K.K.),
and MEXT KAKENHI Grant Nos.~JP15H05889, and
JP16H0877 (K.K.). T.M. thanks the members of the ICG Group at
Portsmouth for their hospitality.

\appendix
%
%
\makeatletter
\renewcommand{\theequation}{A.\arabic{equation}}
\renewcommand{\appendixname}{{\rm Appendix ~}}
\makeatother
 
\section*{Appendix: Explicit form of $\Gamma^I_{\;JK}$}

According to the action (\ref{Eaction}), the metric of the field space ${\cal G}_{IJ}$ and its inverse are given as
\begin{align}
 {\cal G}_{IJ} =                          
  \begin{pmatrix}
   1 &0\\
   0 &e^{-2\alpha\phi}
  \end{pmatrix}
 ,
 & \ \ \ \ \ \
  {\cal G}^{IJ} =                          
  \begin{pmatrix}
   1 &0\\
   0 &e^{2\alpha\phi}
  \end{pmatrix}.
\end{align}
Using the definition of the Christoffel symbols
\begin{equation}
 \Gamma^I_{JK} = \frac{1}{2} {\cal G}^{IL} \left(\partial_J {\cal G}_{LK} + \partial_K{\cal G}_{JL}
					    - \partial_L{\cal G}_{JK}  \right),
\end{equation}
we find the various components to be given as
\begin{align}
 \Gamma^\chi_{\chi\chi} &= 0, \qquad   \Gamma^\chi_{\chi\phi} =  \Gamma^\chi_{\phi\chi} = -\alpha,
  \qquad    \Gamma^\chi_{\phi\phi}=0,\\
 \Gamma^\phi_{\phi\phi} &= 0, \qquad    \Gamma^\phi_{\phi\chi} = \Gamma^\phi_{\chi\phi}= 0,
  \qquad \quad \Gamma^\phi_{\chi\chi}=\alpha e^{-2\alpha\phi}.
\end{align}

%
%

\bibliographystyle{JHEP}
\bibliography{references}

\end{document}